\documentclass[final,3p,times,twocolumn,a4]{elsarticle}

\usepackage{hyperref}

\def\bea{\begin{eqnarray}}
\def\eea{\end{eqnarray}}

\def\sp{\kern +3pt}
\def\sm{\kern -7pt}
\def\spQ{\kern +6pt}

\def\bea{\begin{eqnarray}}
\def\eea{\end{eqnarray}}

\def\be{\begin{equation}}
\def\ee{\end{equation}}
\def\ba{\begin{eqnarray}}
\def\ea{\end{eqnarray}}

\usepackage{graphics}
\usepackage{graphicx}
\usepackage{epsf}
\usepackage{amsmath}
\usepackage{amssymb}
\usepackage{bbold}

\usepackage{xcolor}

\journal{Physics Letters B}

\begin{document}

\begin{frontmatter}



\title{Electromagnetic $|G_E/G_M|$ ratios of hyperons
at large timelike $q^2$}
\author{G.~Ramalho$^1$, M.~T.~Pe\~na$^{2,3}$,  
K.~Tsushima$^4$ and Myung-Ki Cheoun$^1$}
\vspace{-0.1in}

\address{$^1$Department of Physics and OMEG Institute, Soongsil University, \\
Seoul 06978, Republic of Korea} 
\address{
$^2$LIP, Laborat\'orio de Instrumenta\c{c}\~ao e F\'{i}sica 
Experimental de Part\'{i}culas, \\
Avenida Professor Gama Pinto, 1649-003 Lisboa, Portugal}
\address{
$^3$Departamento de F\'{i}sica e Departamento de Engenharia e Ci\^encias Nucleares, \\
Instituto Superior T\'ecnico (IST),
Universidade de Lisboa, 
Avenida Rovisco Pais, 1049-001 Lisboa, Portugal} 
\address{$^4$Laborat\'orio de 
F\'{i}sica Te\'orica e Computacional -- LFTC,
Universidade Cidade de  S\~ao Paulo,  
01506-000,  S\~ao Paulo, SP, Brazil}


\begin{abstract}
In recent years, it has become possible to
measure not only the magnitude of the electric ($G_E$)
and magnetic ($G_M$) form factors of spin $\frac{1}{2}$ baryons,
but also to measure the relative phases
of those quantities in the timelike kinematic region.
Aiming to interpret present $|G_E/G_M|$ data on hyperons of the baryon octet,
as well as to predict future data, we present model calculations of
that ratio for large invariant 4-momentum square $q^2$ in the timelike region ($q^2>0$). 
Without any further parameter fitting,we extend to the timelike region
a covariant quark model previously developed
to describe the kinematic spacelike region ($q^2 \le 0$)
of the baryon octet form factors.
The model takes into account
both the effects of valence quarks and the excitations
of the meson cloud which dresses the baryons.
This application to the timelike region assumes
an approximation based on unitarity and  analyticity that is  valid only
in the large $q^2$ region.
Using the recent data from BESIII we establish the regime
of validity of this approximation.   
We report here that our results for the effective form factor
(combination of $|G_E|$ and $|G_M|$) are in good agreement
with the data already for $q^2$ values above 15 GeV$^2$.
In addition, a more conservative onset of the validity
of the approximation is provided
by the newly available $|G_E/G_M|$ data which suggest that our predictions may 
be compared against data for $q^2 \ge $ 20 GeV$^2$.
This is expected in the near future, when the range
of the present measurements is expanded
to the 20--50 GeV$^2$ region.
\end{abstract}


\end{frontmatter}

\section{Introduction}
\label{secIntro}

In recent years it became possible to experimentally probe the structure
of short-lived baryons through
$e^+ e^- \to B \bar B$ and $p \bar p \to B \bar B$ experiments in
the timelike kinematic region above the production threshold
$q^2 \ge  4 M_B^2 $ ($M_B$ is the mass of the
baryon)~\cite{BESIII15a,BESIII20a,Dobbs17a,BESIII19a,BESIII23a,HyperonTL}.
These timelike studies complement the two last decades knowledge
on the spacelike region\footnote{In this work,  for simplicity,  we include into the spacelike region the photon point $q^2 = 0$,
due to the physical continuity between the regions $q^2 < 0$ and $q^2=0$.}
($q^2 \le 0$) based on results from electron scattering experiments~\cite{NSTAR,Aznauryan12a,PPNP2023}.

The electromagnetic structure of the baryons is represented
by spin ($J$) and parity ($P$)
dependent structure form factors.
In the timelike region these form factors are complex functions of the 
$e^+ e^-$ or $p \bar p$ square center-of-mass energy 
$s= q^2$~\cite{PPNP2023,Pacetti15a,Denig13,OmegaFF,Seth13,Cabibbo61a,DM2}.
Of particular interest are the spin $\frac{1}{2}$ baryons ($J^P =\frac{1}{2}^\pm$),
including the nucleon and hyperons 
characterized by two form factors only, the electric and magnetic form factors.
We focus here on hyperons because the experimental information
about the nucleon is at the moment less complete.
Also, our model uncertainty decreases with increasing $q^2$,
and the convergence of the nucleon data to the asymptotic large $q^2$
behavior is expected to emerge at much larger $q^2$ values, due
the dominance of light quark dynamics.

Given the short lifetime of hyperons, the challenges in the determination
of their electromagnetic properties
in the timelike region, where their weak decays can be analyzed
in detail~\cite{BESIII20a}
are less significant than in the spacelike region.

The first experiments measured the total integrated cross section
of the $e^+ e^- \to B \bar B$ reactions for spin $\frac{1}{2}$ baryons,
which determine the magnitude of the  "effective"
form factor $G$ defined as~\cite{BESIII20a,HyperonTL}
\ba
|G (q^2)|^2 = \frac{2 \tau |G_M (q^2)|^2 + |G_E (q^2)|^2}{2 \tau + 1},
\label{eqGeff}
\ea
where $\tau = \frac{q^2}{4M_B^2}$.

These experiments provided information only about
a combination of the electric and magnetic form factors,
and did not reveal their relative phases.
The separate determination of $|G_E|$ and $|G_M|$ or of the ratio
$|G_E/G_M|$ for spin $\frac{1}{2}$ baryons in the timelike region 
can be obtained by further measuring  the differential
cross section~\cite{BESIII19a,BESIII23a,BESIII22a}, while
the knowledge of the relative phase between the two form
factors demands in addition the determination of polarization cross sections
(which involve complete spin separation)~\cite{BESIII20a,BESIII19a,BESIII23a}.
Today, indeed it is possible to measure the ratio $|G_E/G_M|$
and the relative angle between the form factors
for a few baryons~\cite{BESIII19a,BESIII23a,Aubert07a}.
Most measurements to date have been done
for ground state hyperons, but in the future excited states
may be accessed. In a first experimental period the proton and the neutron form factors
were extracted under some assumptions about
the relation between $G_E$ and $G_M$~\cite{Pacetti15a,Denig13,Aubert07a}.

Although the initial measurements were performed near the threshold,
it is expected that with increased accelerator power,
the experimental energy will go higher up to the 20--50 GeV$^2$
range. This extension will allow us to test the predictions described in this work.

Measurements of the effective elastic form factors
of the $\Lambda$, $\Sigma^{0,\pm}$, $\Xi^{0,-}$
as well as the transition form factors $\Sigma^0 \to \Lambda$
have been performed at BaBar~\cite{Aubert07a}, CLEO~\cite{Dobbs17a,Dobbs14a},
Belle~\cite{Belle23a} and BESIII~\cite{BESIII19a,BESIII23a,BESIII22a,BESIII23d,BESIII21a,BESIII23x,BESIII18b,BESIII21d,BESIII24a,BESIII24b,BESIII22X,BESIII21b,BESIII20c,BESIII23b,BESIII23c}.
Measurements of the ratios  $|G_E/G_M|$ for the $\Lambda$ and $\Sigma^+$
have also been determined recently at BESIII~\cite{BESIII19a,BESIII23a,BESIII22a}.
Complete experiments, that measure  $|G_E|$, $|G_M|$, 
and the relative phase between $G_E$ and $G_M$ have been performed
for the $\Lambda$ and $\Sigma^+$ at BESIII~\cite{BESIII19a,BESIII23a}, 
and even more measurements are planned for a near future at BESIII
and PANDA~\cite{PANDA16a,Schonning20a}.

In this work, we enlarge a previous theoretical study of the effective form factors
of $J^P= \frac{1}{2}^+$ hyperons~\cite{HyperonTL} to calculate the $|G_E/G_M|$ ratios 
for large $q^2$.
Calculations in this momentum transfer region are important,
given that at very large $q^2$, $|G_M (q^2)|$ dominates
the effective form factor $|G (q^2)|$, allowing us to probe where that regime sets in.

The numerical calculations are based on the
covariant spectator quark model formalism 
for the spacelike region~\cite{NSTAR,Nucleon,Omega,NSTAR2017}
extended to the timelike region
using asymptotic relations valid for large $q^2$~\cite{HyperonTL}.
The model has been used in the study
of electromagnetic structure of 
nucleon resonances~\cite{NSTAR2017,NDelta,Roper,N1535,N1520,Lattice}
and other baryons
systems~\cite{Baryons2,Octet13,Octet2}.

In the case of the nucleon is it possible to
compare the magnitude of $G_E/G_M$ in the spacelike
and timelike regions~\cite{Seth13,Nucleon,Octet2,NucleonGEGM,Puckett17,Gustafsson01a}.
Timelike data on the nucleon electromagnetic
form factors can be found
in Refs.~\cite{BESIII15a,BESIII20b,BESIII-proton,BaBar13a,Aubert06,Fermilab}
for the proton and in Refs.~\cite{DM2,BESIII-Neutron1,BESIII-Neutron2,SND-Neutron,Antonelli98} for the neutron.
Theoretical studies of the nucleon electromagnetic form factors
in the timelike region can be found in
Refs.~\cite{Yan24a,Lin21a,Lin22b,Kuraev12,Gustafsson21a,Lorenz15a,Bianconi15a,Cao22b,Yang24a}.
At the moment the measurements of the nucleon form factors in
the timelike region are incomplete, since their relative phases
have not been measured.
The data of the proton and neutron effective form factors
revealed an oscillatory dependence on the center-of-mass
energy~\cite{BESIII20a,BESIII-proton,BESIII-Neutron1,Lorenz15a,Bianconi15a}.
Different explanations have been proposed~\cite{Yan24a,Lin22b,Cao22b,Gustafsson22a}.   
In the present work we do not discuss this subject because
we are focused on the asymptotic region, where
the oscillatory component is expected to be suppressed.

Beyond the members of the baryon octet, there are also measurements
of the effective form factors of the $\Omega^-$, $\Delta$
and $\Lambda_c^+$~\cite{Dobbs17a,Dobbs14a,BESIII18a,BESIII23t,BESIII23z}.
Theoretical works on hyperon form factors
in the timelike region can be found in 
Refs.~\cite{Gustafsson22a,Korner77,Haidenbauer92,Haidenbauer16,Lin22a,Yang19,Cao18,Mangoni21a,Bianconi22a,Dai23a,Dai22a,Bystritskiy22a}.

\section{Formalism}
\label{secFormalism}

Within the one-photon-exchange approximation,~\cite{BESIII15a,BESIII20a}
we can write the $e^+ e^- \to B \bar B$ differential
cross section, where $B$ stands for
a spin $\frac{1}{2}$ baryon $B\left( \frac{1}{2}^\pm \right)$, as
\ba
& & 
\frac{d \sigma_{\rm Born}}{d \Omega} (q^2) =
\label{eqDiffCX}
\\
& &  \frac{\alpha^2 \beta C}{4 q^2}
  \left[ |G_M (q^2)|^2 (1 + \cos^2 \theta)
    + \frac{1}{\tau} |G_E (q^2)|^2 \sin^2 \theta \right],
  \nonumber
\ea
where $\theta$ is the angle between the baryon
and the initial photon in the center-of-mass frame,
$\alpha \simeq 1/137$ is the fine structure constant,
$\beta= \sqrt{1 - \frac{4M_B^2}{q^2}}$ is a kinematic factor
(the baryon speed in $c=1$ units) and 
$C$ is the Sommerfeld-Gamow factor.
This  factor takes into account the Coulomb interaction
and can be written as $C = \frac{y}{1-\exp(-y)}$
with $y= \frac{\pi \alpha}{\beta} \frac{2 M_B}{\sqrt{q^2}}$
for charged particles, and $C=1$ for 
neutral particles~\cite{Pacetti15a,Aubert06}.
For charged particles the factor $C$ converges rapidly to $C \simeq 1$ 
when $q^2$ increases.

By integrating the previous relation, we obtain
\ba
\sigma_{\rm Born} (q^2) =
\frac{4 \pi \alpha^2 \beta C}{3 q^2}
\left( 1 + \frac{1}{2 \tau} \right) |G (q^2)|^2,
\label{eqCX}
\ea
using Eq.~(\ref{eqGeff}).
We can conclude then that the effective form
factor can be calculated directly from the integrated cross section.
However, it does not give individual information on  $G_E$ or $G_M$.

To untangle information on the two form factors $G_E$ and $G_M$
one uses  the differential cross section (\ref{eqDiffCX}) by
determining the coefficients of the angular functions
$1 + \cos^2 \theta$ and $\sin^2 \theta$~\cite{BESIII15a,BESIII20a,BESIII21d,BESIII23b}.
Combining the two results, one can extract the ratio $|G_E/G_M|$.
The measurement of the phase between the form factors
requires in addition the measurement of cross section
with defined polarizations of the baryons~\cite{BESIII23b}.


Finally, at threshold, where $q^2= 4 M_B^2$, the functions
$G_E$ and $G_M$ are such that $G_E = G_M$,
as a consequence of the definition
of these Sachs from factors in terms of the kinematic independent
Dirac and Pauli form factors. Therefore one expects that $G_E/G_M \simeq 1$
for measurements 
near the threshold, and the non-zero phases between the form factors are to be sought 
above the threshold  only ($q^2 > 4 M_B^2$)~\cite{BESIII20a}.

The main conclusion of this section is that in the case
of a spin $\frac{1}{2}$ baryon, to obtain 
the absolute values of the two form factors from
experiments of the  differential cross section
it is necessary to measure accurately angular distributions
of the differential cross section
besides the effective form factor $|G(q^2)|$.
However, as we will see in the next section the range
for non-zero imaginary parts cannot extend to indefinitely large $|q^2|$.

\section{Calculation of electromagnetic form factors at large $q^2$}
\label{secLarge-q2}

In this work the electromagnetic form factors $G_E$ and $G_M$
in the timelike region  are obtained from extrapolating the results
from the region $q^2=-Q^2 \le  0$ to the region $q^2 > 0$.

For that purpose we consider the large-$|q^2|$ relations 
\ba
G_\ell (q^2)= G_\ell (-q^2)
\label{eqAssmp1}
\ea
for $\ell=E,M$. 
These asymptotic relations are a consequence of
two general mathematical principles: 
unitarity as well as the Phragm\'en-Lindel\"{o}f theorem,
which is valid for analytic functions~\cite{Pacetti15a}.

We notice that this theorem implies that the imaginary parts of
the form factors in the timelike region 
must be negligible for large $q^2$,
since the form factors are real in the spacelike region.

The relations (\ref{eqAssmp1}) are strictly valid in the
mathematical limit $q^2 \to +\infty$.
Alone the relations provide no information on the
range where approximated results can be expected to have
a certain accuracy.
But the large physical $q^2$ scale where the mathematical
theorem starts to be valid can be determined by the comparison
with the physical data, which we do here.


\begin{figure*}[t]
\vspace{.5cm}
\centerline{
\mbox{
\includegraphics[width=2.2in]{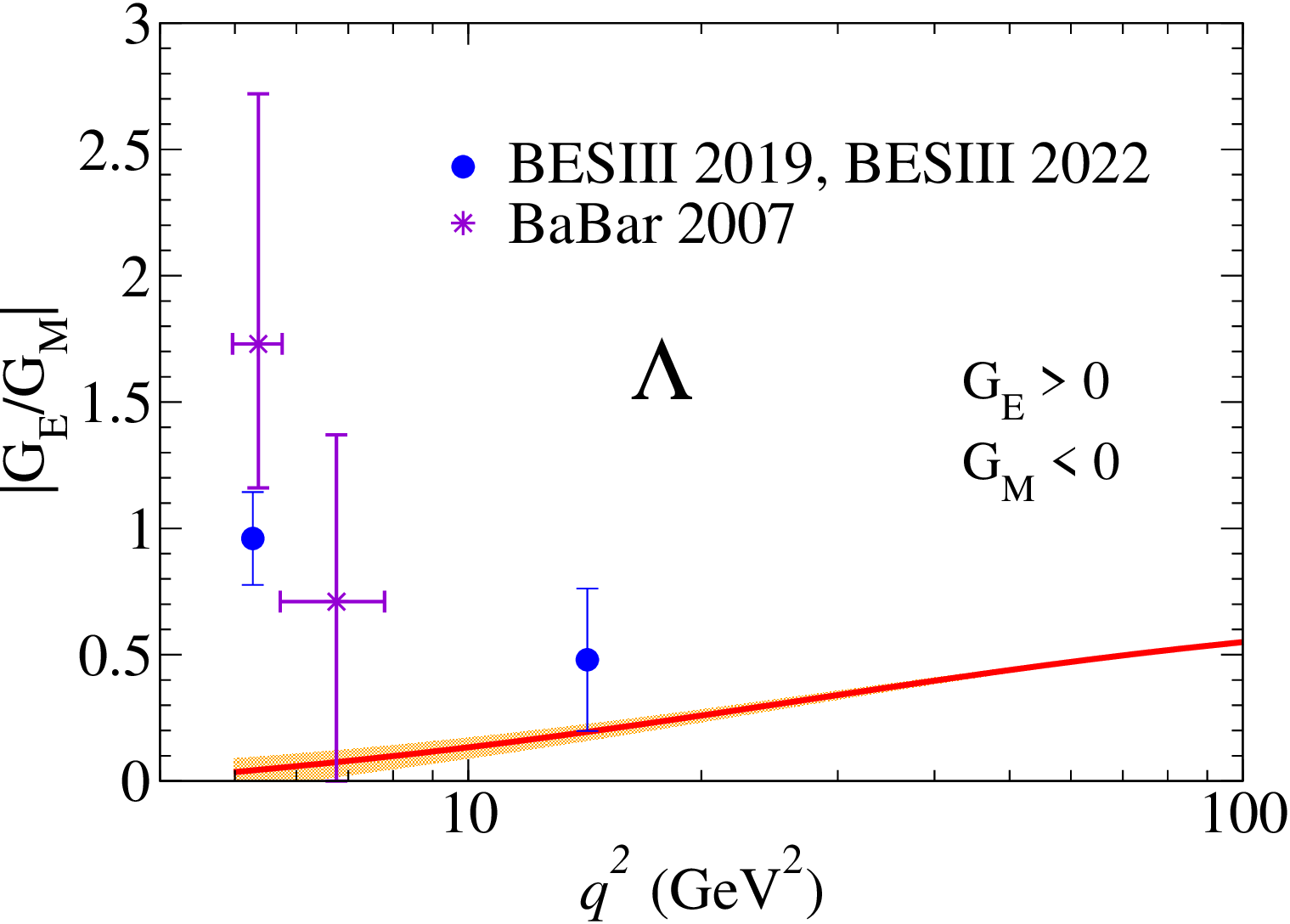} \hspace{.1cm}
\includegraphics[width=2.2in]{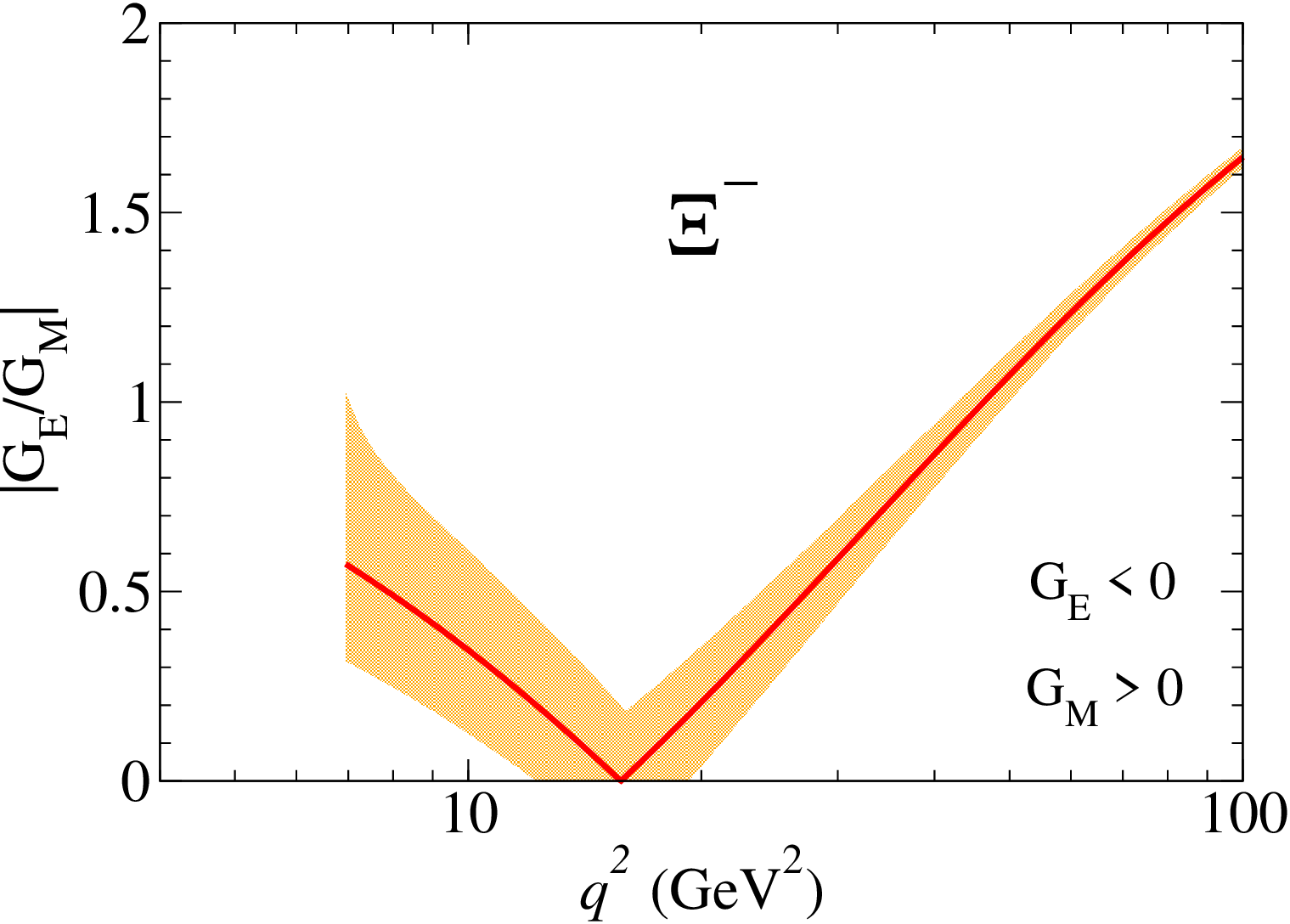}
\hspace{.1cm}
\includegraphics[width=2.2in]{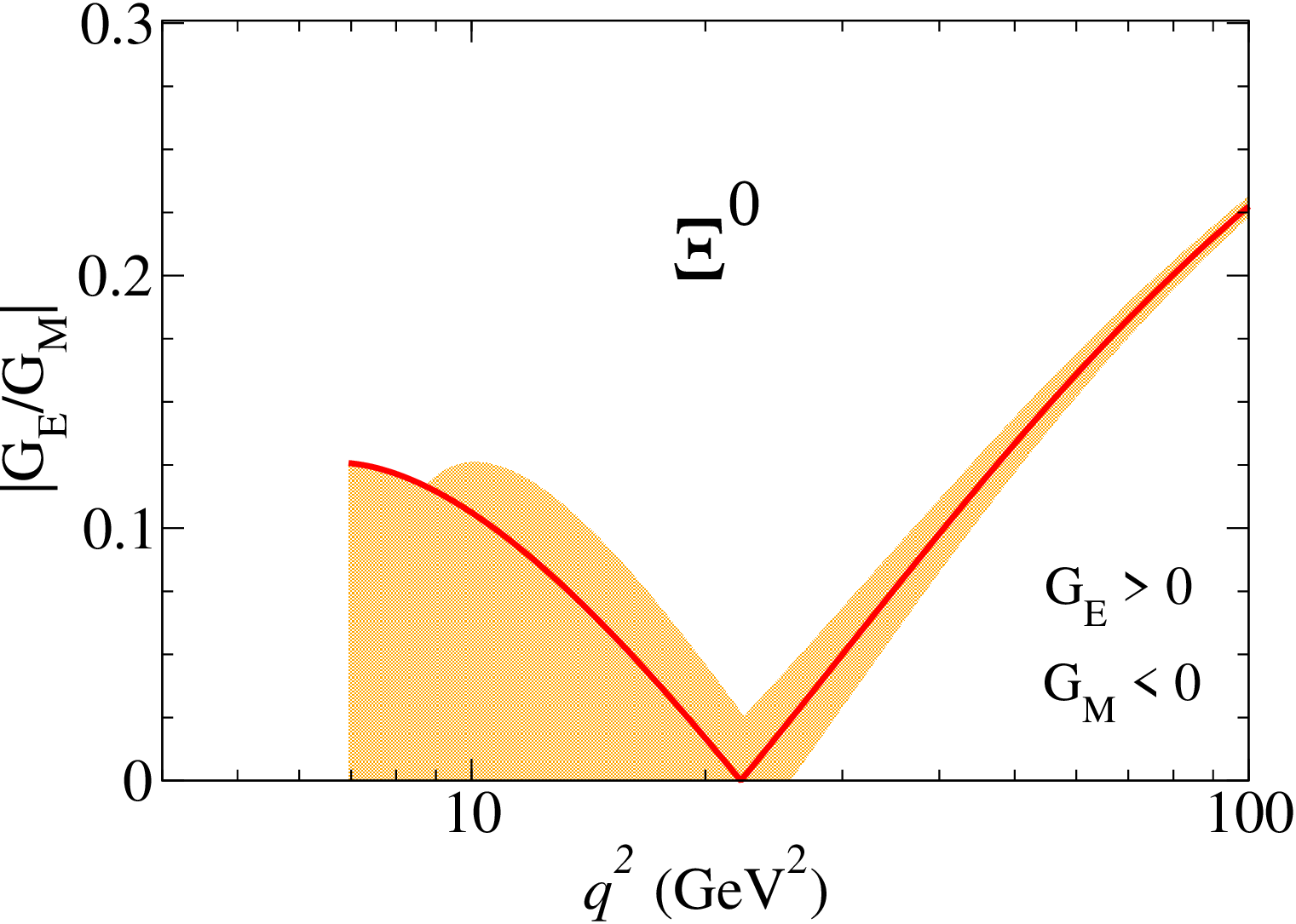} }}
\caption{\footnotesize{
Ratios $|G_E/G_M|$ for $\Lambda$, $\Xi^0$ and $\Xi^-$.
The data for $\Lambda$ are from BaBar~\cite{Aubert07a} and
BESIII~\cite{BESIII19a,BESIII22a}.}}
\label{figure2}
\end{figure*}

\begin{figure*}[t]
\vspace{.2cm}
\centerline{
\mbox{
\includegraphics[width=2.2in]{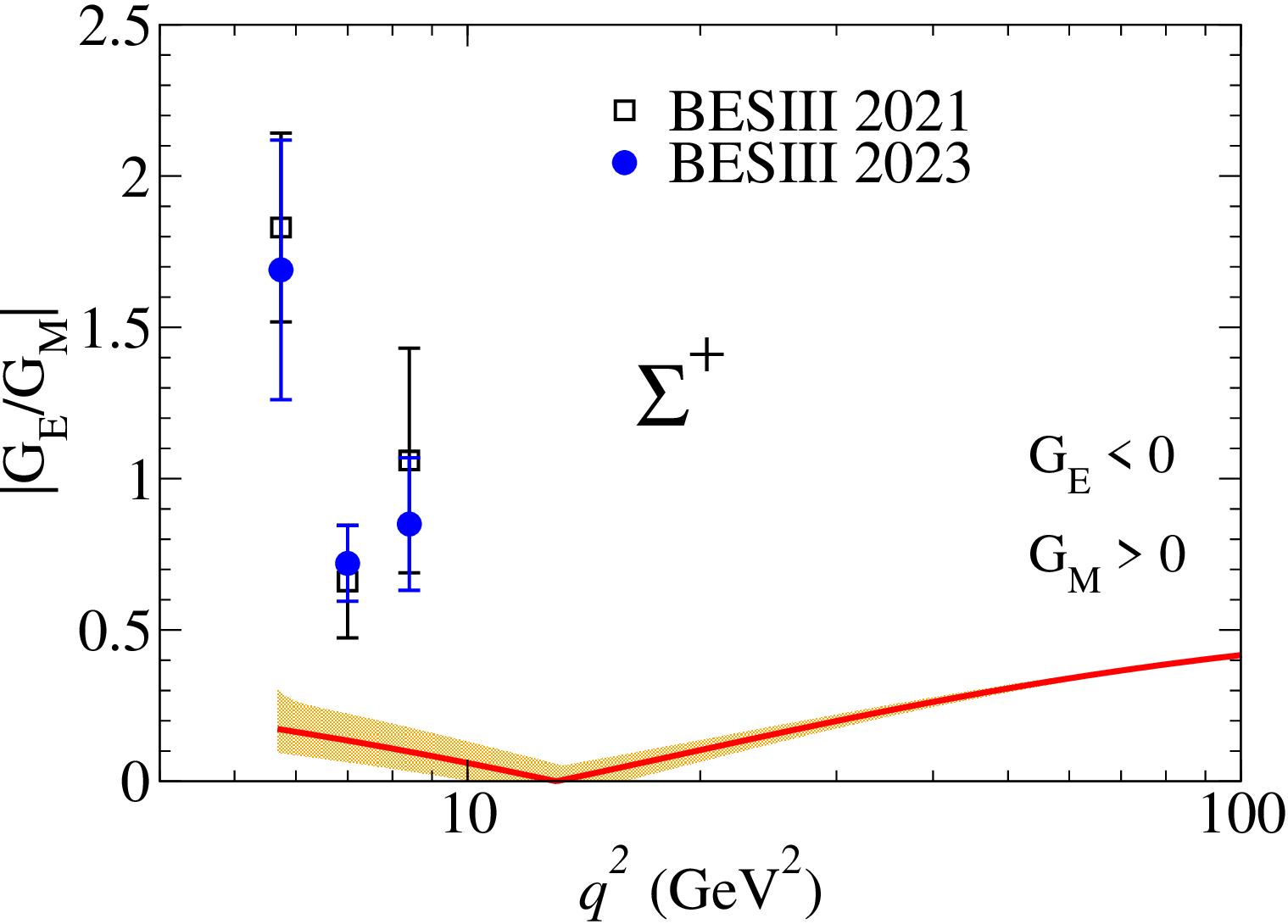} \hspace{.1cm}
\includegraphics[width=2.2in]{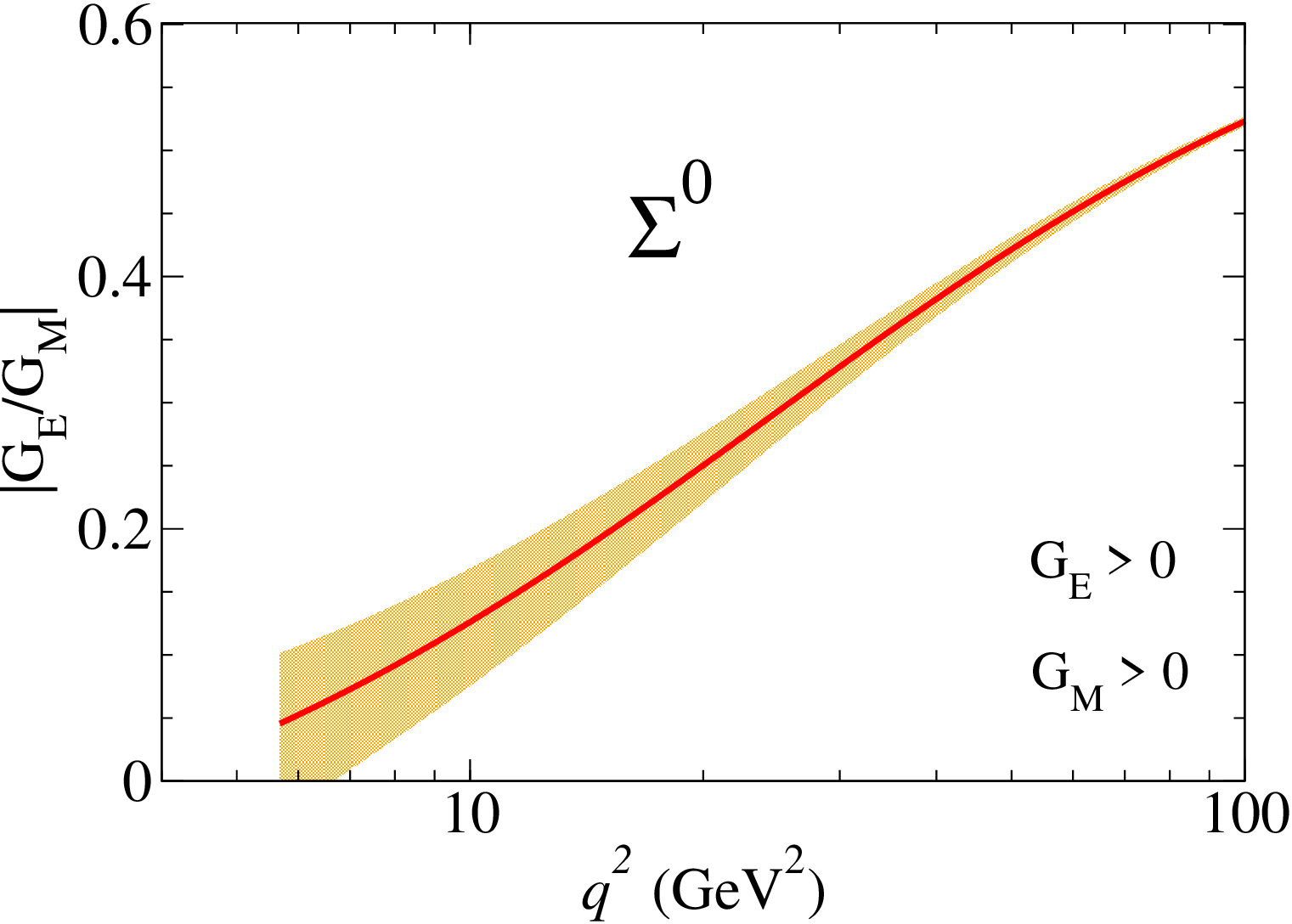}
\hspace{.1cm}
\includegraphics[width=2.2in]{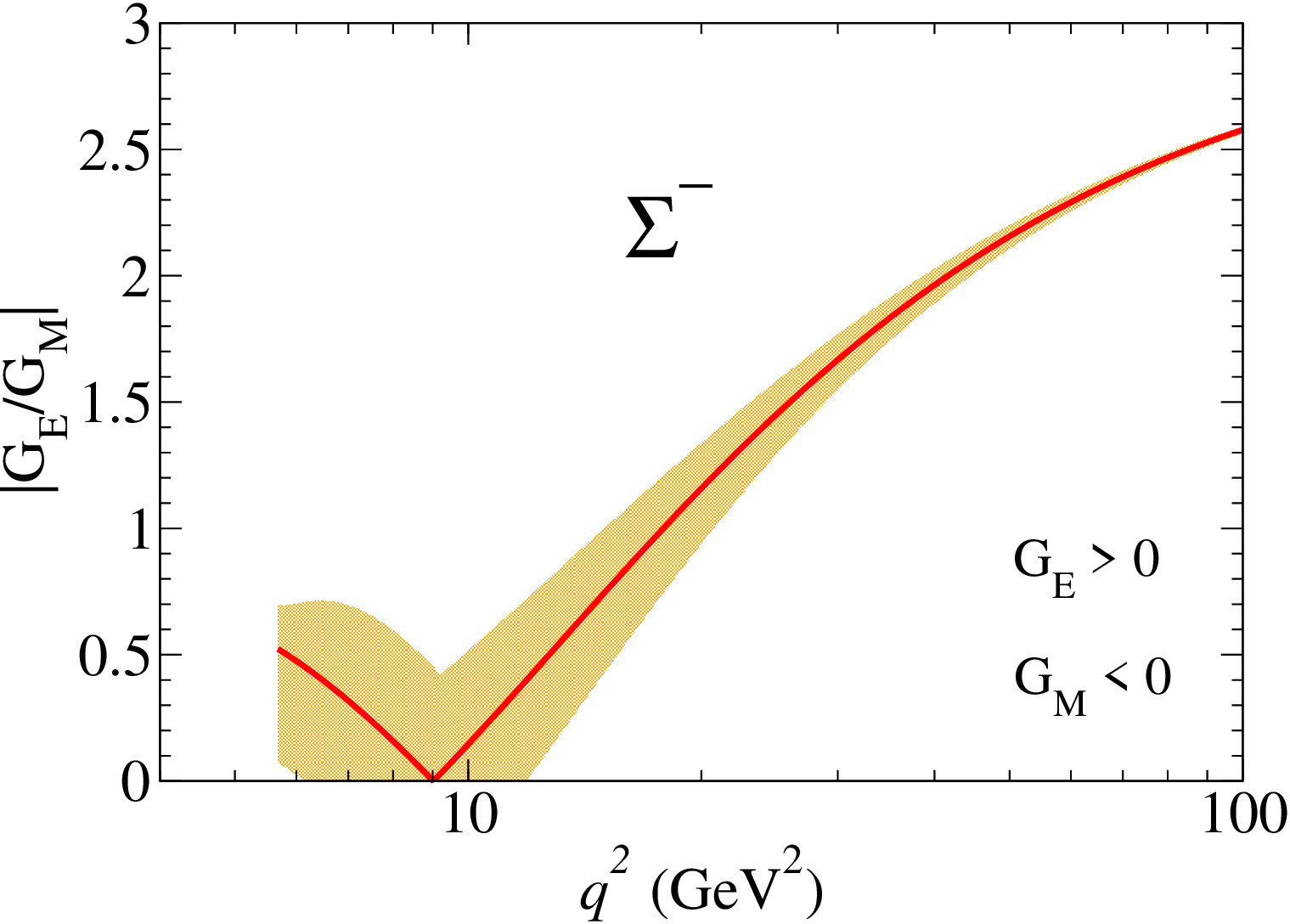} }}
\caption{\footnotesize{Ratios $|G_E/G_M|$ for $\Sigma^+$, $\Sigma^0$ and  $\Sigma^-$.
The data for the $\Sigma^+$ are from BESIII~\cite{BESIII23a,BESIII21d}.}}
\label{figure1}
\end{figure*}

\begin{figure*}[t]
\centerline{
\mbox{
\includegraphics[width=2.8in]{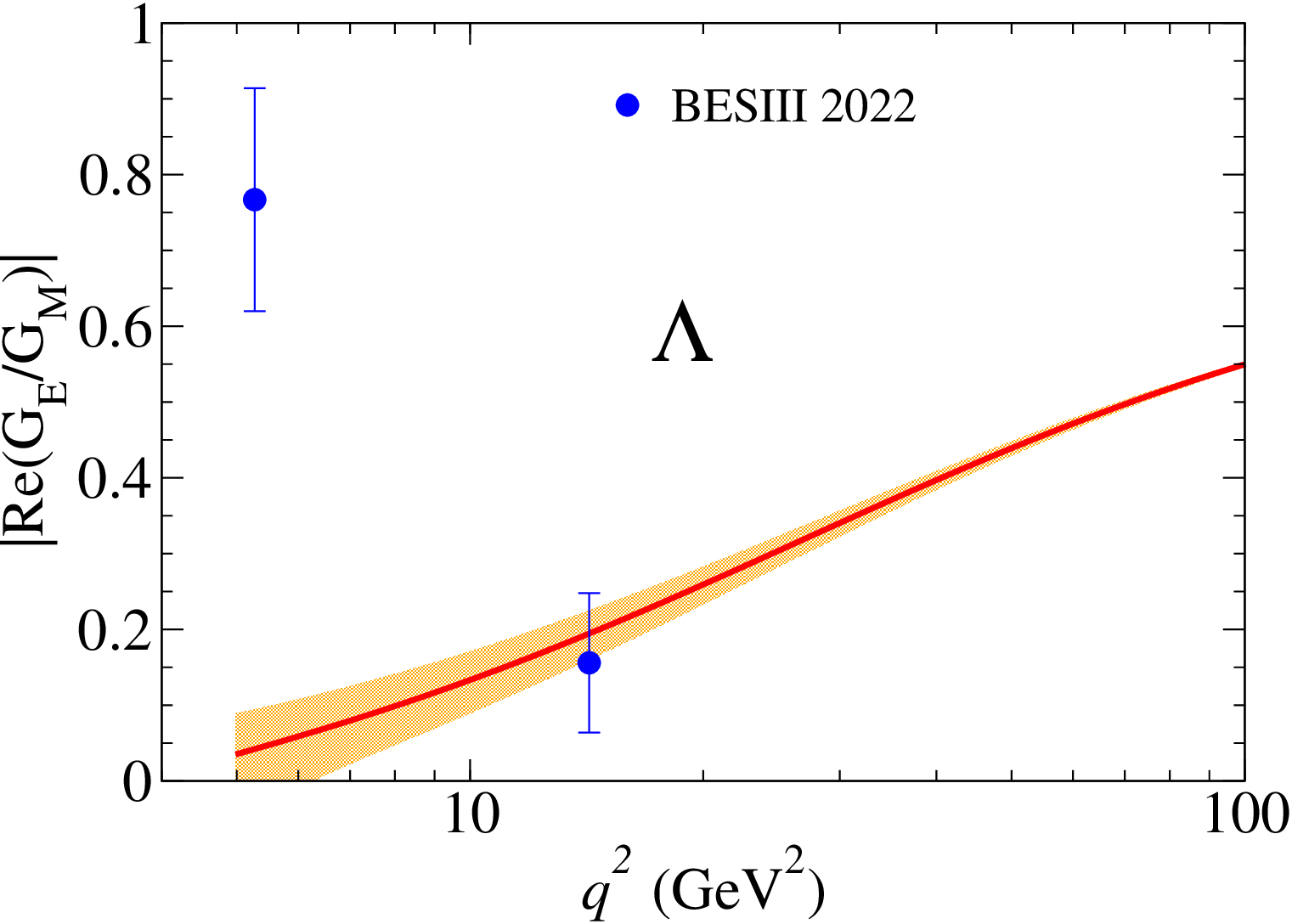} \hspace{1.2cm}
\includegraphics[width=2.8in]{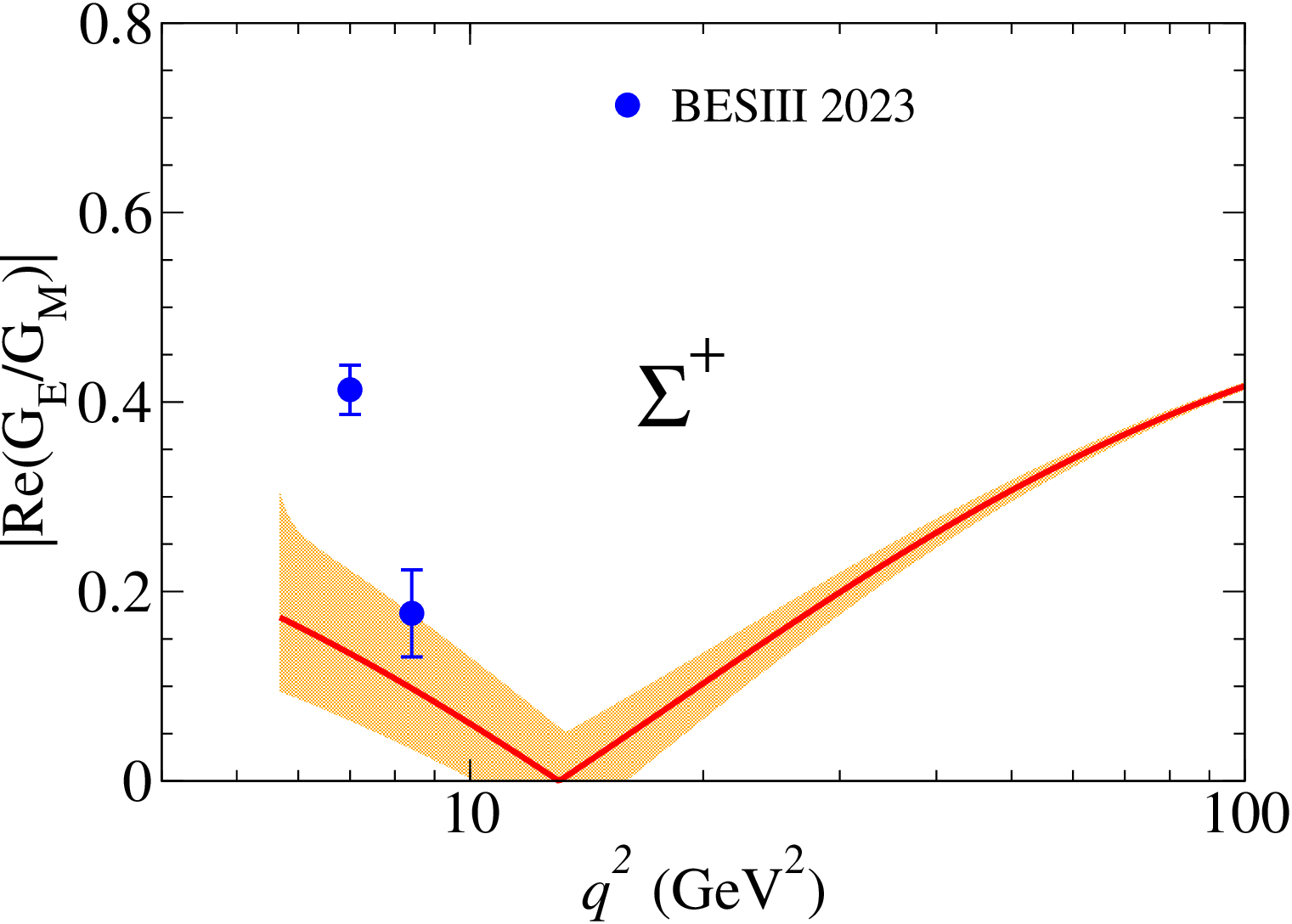} }}
\caption{\footnotesize{Model calculations
    of $|G_E/G_M|$ for $\Lambda$ and $\Sigma^+$
    compared with the $|{\rm Re}(G_E/G_M)|$ data from BESIII 
    for $\Lambda$~\cite{BESIII22a} and $\Sigma^+$~\cite{BESIII23a}.}}
\label{figure3}
\end{figure*}

Given the character of approximation,
based on asymptotic relations,
we expect our estimates here to be accurate only for large $q^2$,
deviating  from the condition $G_E = G_M$ at the threshold.

Since the relations (\ref{eqAssmp1}) are valid for
the large-$|q^2|$, and we want to make predictions
for finite $q^2$, we consider corrections to these
asymptotic relations.
We start by noticing that the elastic form factors
are divided into two regions of  $q^2$: $(- \infty, 0]$ (spacelike region)
and  $[4M_B^2, + \infty)$,  with a gap of $4M_B^2$ length
between the photon point ($q^2=0$) and
the threshold of the baryon-antibaryon production ($q^2 = 4 M_B^2$).  
The interval $(0,4M_B^2)$ defines an unphysical region
where no physical baryons can be detected~\cite{Pacetti15a,Denig13},
but where the peak structure in the form factors can be inferred
from the $e^+ e^- \to \pi^+ \pi^-$ cross sections.
Only for much larger values of $q^2$
above the threshold ($q^2= 4 M_B^2$), we expect that
the tail of the form factors becomes symmetric to the 
smooth one of the spacelike region ($q^2 \le 0$).

It is then clear that $q^2=0$ is not the center
of reflection of the asymptotic symmetry relation~(\ref{eqAssmp1}).
But there is an ambiguity about the exact location of this center, 
and to take into account this ambiguity we replace
$q^2 \to q^2 \left( 1 - \frac{2M_B^2}{q^2} \right)$ in~(\ref{eqAssmp1}), 
which introduces finite corrections ($-2M_B^2$)
to the limit $q^2= + \infty$.
Therefore we use~\cite{HyperonTL}
\ba
G_\ell (q^2)= G_\ell^{\rm SL} (q^2 - 2 M_B^2)
\label{eqAssmp2}
\ea
for $\ell =E,M$. 
The label SL on the r.h.s.~indicates that 
the $G_\ell^{\rm SL} (Q^2)$ is calculated
in the spacelike region ($Q^2 = -q^2 \ge 0$).
The expression (\ref{eqAssmp2}) estimates the reflection
symmetry point underlying (\ref{eqAssmp1})
exactly at the center of the unphysical interval $(0,4M_B^2)$.

In summary, the measured form factors in the timelike region are complex functions
characterized by their magnitudes, $|G_E|$, $|G_M|$, and their different phases.
Our model calculation is based on results obtained in the $q^2 \le 0$
region and the application of the large-$q^2$ relation~(\ref{eqAssmp2}).
This way, by construction, we obtain real numbers for $G_E$
and $G_M$ that carry $\pm$ signs.
In principle, for very large $q^2$ the comparison of our results with the
experimental values for $G_E/G_M$ is possible
since in that region the relative phase can only have two possibilities,
$\Delta \Phi =0$ (when the form factors have the same sign)
or  $\Delta \Phi = \pi$ (if they have different signs).

For a control of the approximation we explored the 
band of variation of the form factors with respect to a reasonable variation
of the symmetry point around this central value:
when the functions $G_\ell (q^2)$ are positive (negative), since
they are expected to monotonically decrease (increase)
with increasing $q^2$, we take
the upper (lower) limit values to be $G_\ell (q^2)= G_\ell^{\rm SL} (q^2 - 4M_B^2)$
and the lower (upper) limit  $G_\ell (q^2)= G_\ell^{\rm SL} (q^2)$.
This procedure provides a band of variation of the
finite corrections to the asymptotic symmetry identity.
When $q^2$ is very large, the variation of the results becomes
very narrow and the three estimates (central, upper and lower limits)
converge to the same value, providing accurate predictions
that can be tested by experimental data.

The relations (\ref{eqAssmp2}) are naturally valid
in the perturbative QCD (pQCD) region, where the
form factors are regulated by power laws~\cite{Brodsky75a,Mergell96a}.
Is it possible, however, that they start to be observed earlier than the pQCD regime. 
The origin of the symmetry relations is unitarity and analyticity.
To begin with, the hadron bound states and decays
of the hadrons on meson-baryon states
seen as peaked structures of the timelike form factors
are limited to the confinement region (lower energy).
The comparison with data provides information on the onset
of the regime of the asymptotic reflection symmetry (\ref{eqAssmp1}).
This is further discussed in Section~\ref{secEFF}
on the effective form factors.

\section{Numerical results for $|G_E/G_M|$
  \label{secResults}}

In this section we present our results of the ratio $|G_E/G_M|$ for large $q^2$ values.
The calculation method of the form factors $G_E$ and $G_M$
for the baryon octet family in the timelike region is discussed
in detail in Ref.~\cite{HyperonTL} which uses the
covariant spectator quark model formalism
for the spacelike region~\cite{NSTAR,Nucleon,Omega,NSTAR2017}. 
In its nutshell, we use impulse approximation for the electromagnetic
quark current, which is justified for large $Q^2$, since then
the interacting photon has sufficient resolution to interact with a single quark at a time.
As a result we can integrate over the relative internal variables
of the spectator quarks  and therefore we end up with the
baryon three-quark system regarded 
as a quark-diquark configuration.
In the model, the SU(3) symmetry quark flavor symmetry
is broken at the level of the baryon radial wave functions
and the constituent quark current~\cite{Omega,Octet13,OctetDecuplet12},
and gluon and quark-antiquark dressing
are effectively taken into the constituent quark structure.

The different baryon radial wave functions are parametrized
by two parameters, a short range scale and a long range scale,
consistent with the expected size of the systems
(baryons with strange quarks are more
compact than baryons with light quarks)~\cite{Octet13}.
The quark current is parametrized by
vector meson dominance regulated by
light ($\rho$, $\omega$) and intermediate ($\phi$)
vector meson mass poles~\cite{Omega,Octet13}.

In particular, we use the parameterization
of the baryon octet from Refs.~\cite{Octet13}.
To the valence quark contributions
we add also an effective description of meson cloud
dressing processes~\cite{Lattice,OctetDecuplet12,Timelike2,OctetDecuplet34,LambdaStar},    
which are inevitably present in the low-$Q^2$ region.
In the large-$|Q^2|$ region, the meson cloud contributions are suppressed
and their effects are manifest on the rescaling of
the bare contributions to the form factors due to
the normalization of the hyperon wave functions~\cite{HyperonTL,Octet13}.
This formalism has been used to successfully describe the spacelike electromagnetic
structure data of baryons including the low-lying nucleon resonances~\cite{PPNP2023,NSTAR2017}.
The free parameters of the model are fixed by the
data for the nucleon, baryon octet and
baryon decuplet~\cite{Nucleon,Octet13,Octet2,OctetDecuplet12,OctetDecuplet34,LambdaStar}.

Our calculations of the ratio $|G_E/G_M|$ and their comparison with the data
are shown in Figs.~\ref{figure2} and \ref{figure1},
that depict the results for $\Lambda$, $\Xi^{-,0}$ and  the $\Sigma^{0,\pm}$, respectively.
In the graphs for the $\Lambda$ and $\Sigma^+$ we
include the available data for $|G_E/G_M|$.
How we estimate the uncertainty of the results of our model for each form factor
was defined in the previous section.
The bands of variation for the magnitude of $G_E/G_M$ are determined
from the independent band intervals of variation for the $G_E$ and $G_M$ functions.

The results and their uncertainty bands near the threshold provide only 
qualitative estimates since our results are expected to be valid only for large $q^2$,
as discussed in the previous section.
The deviation from the result  $|G_E/G_M|=1$ at threshold
is a consequence of the approximation used.
The signs of our results for the form factors in the asymptotic
large $q^2$ values are indicated on the
right side in Figs.~\ref{figure2} and \ref{figure1}.

The separated calculations of the form factors $G_E$ and $G_M$
for $\Lambda$, $\Sigma^{0,\pm}$  and $\Xi^{-,0}$ are presented
in the Supplementary Material~\cite{SuppM}.

Note that we are not attempting to predict the zeros
of $G_E/G_M$, rather, we are focused on determining
its magnitude at sufficiently large $q^2$.
In the current formalism, the zeros can only be determined within
an uncertainty band of width $4 M_B^2$ for $q^2$.


\begin{figure*}[t]
\centering

\centerline{
\mbox{
\includegraphics[width=2.8in]{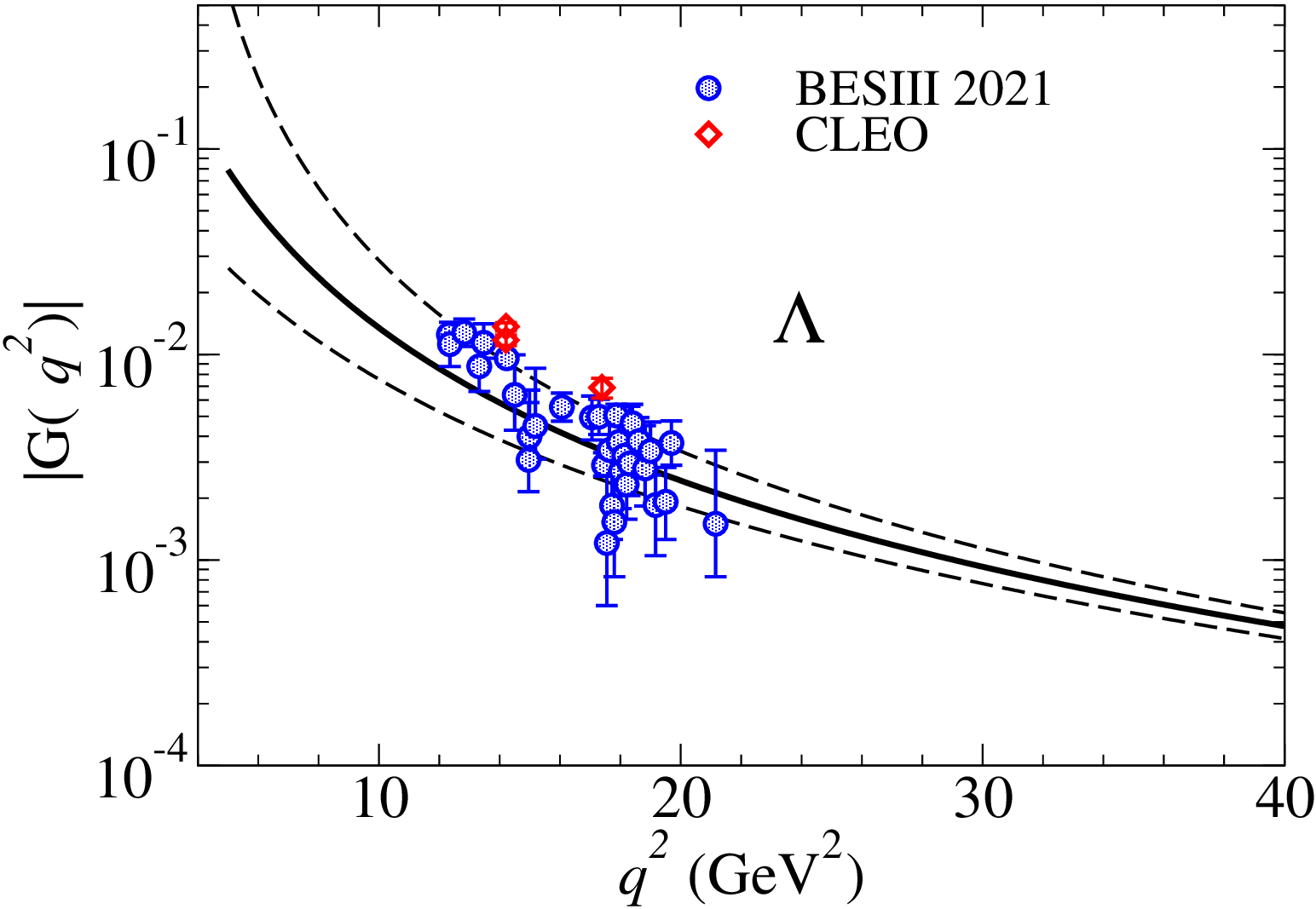} \hspace{1.cm}
\includegraphics[width=2.8in]{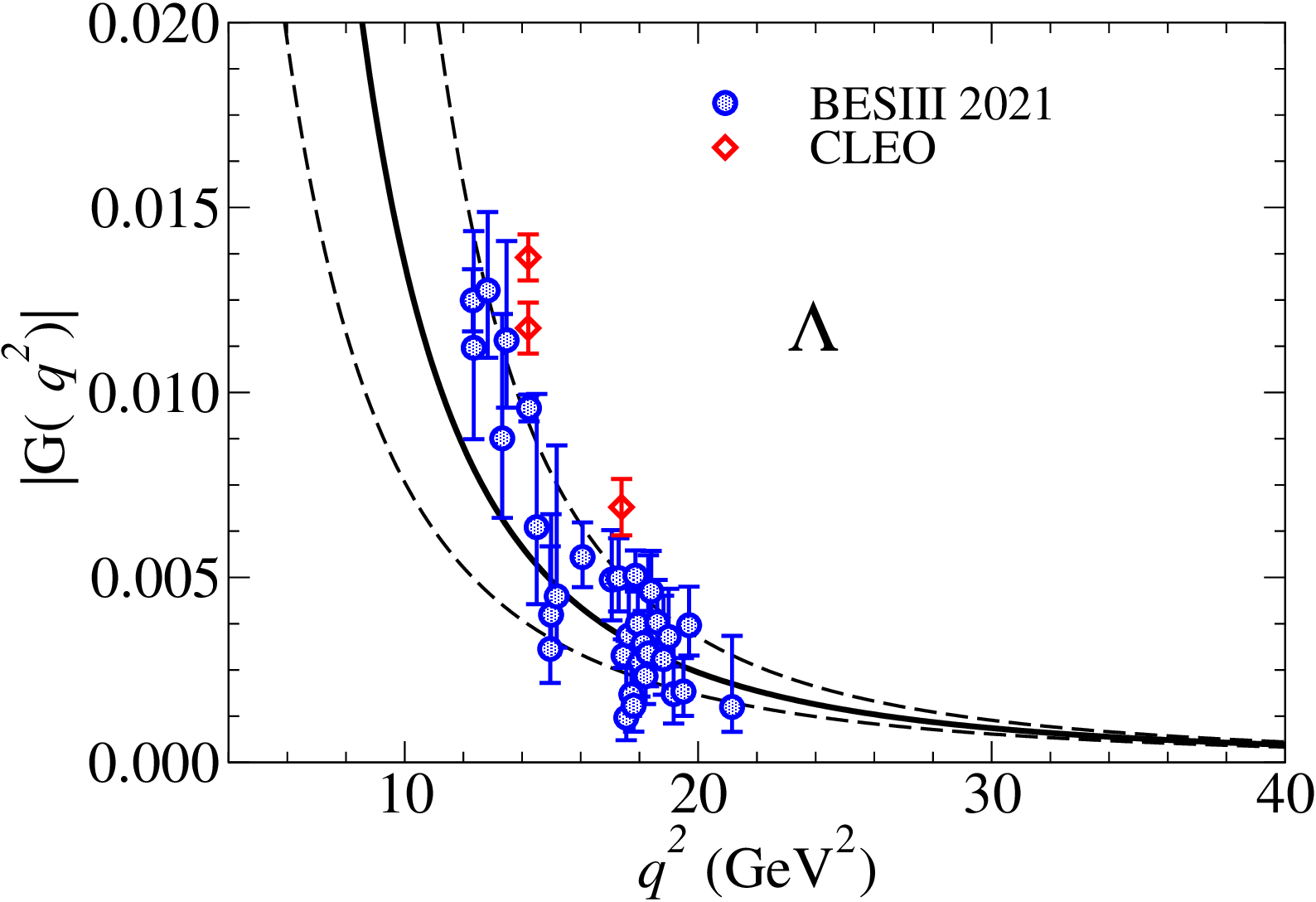} }}
\caption{\footnotesize{Effective form factor $|G(q^2)|$ for $\Lambda$.
    The data are from BaBar~\cite{Aubert07a}, 
    CLEO~\cite{Dobbs17a,Dobbs14a} and BESIII 2021~\cite{BESIII21a}.} \label{figLambda2}}

\bigskip
\vspace{.2cm}

\centerline{
\mbox{
\includegraphics[width=2.8in]{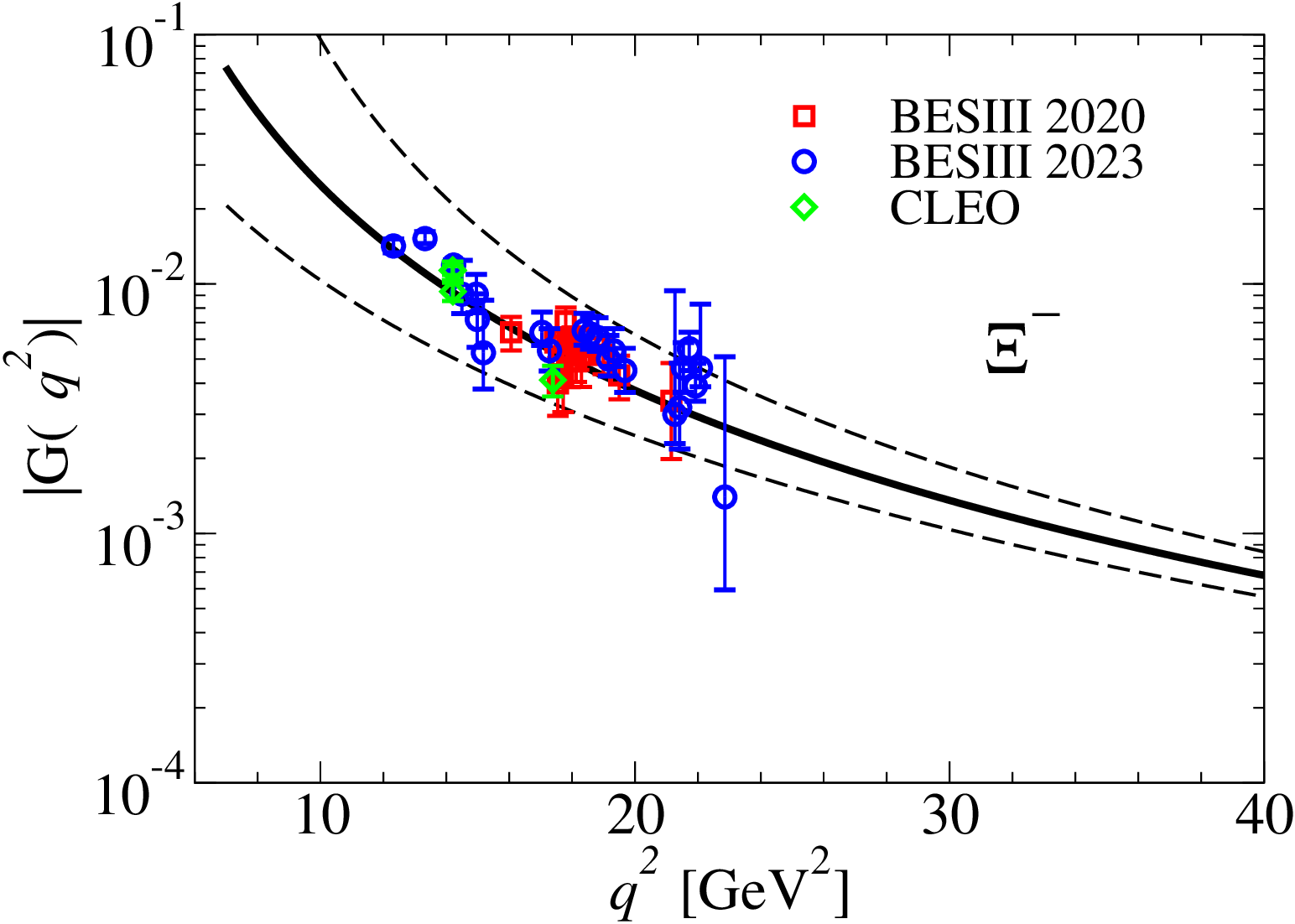} \hspace{1.cm}
\includegraphics[width=2.8in]{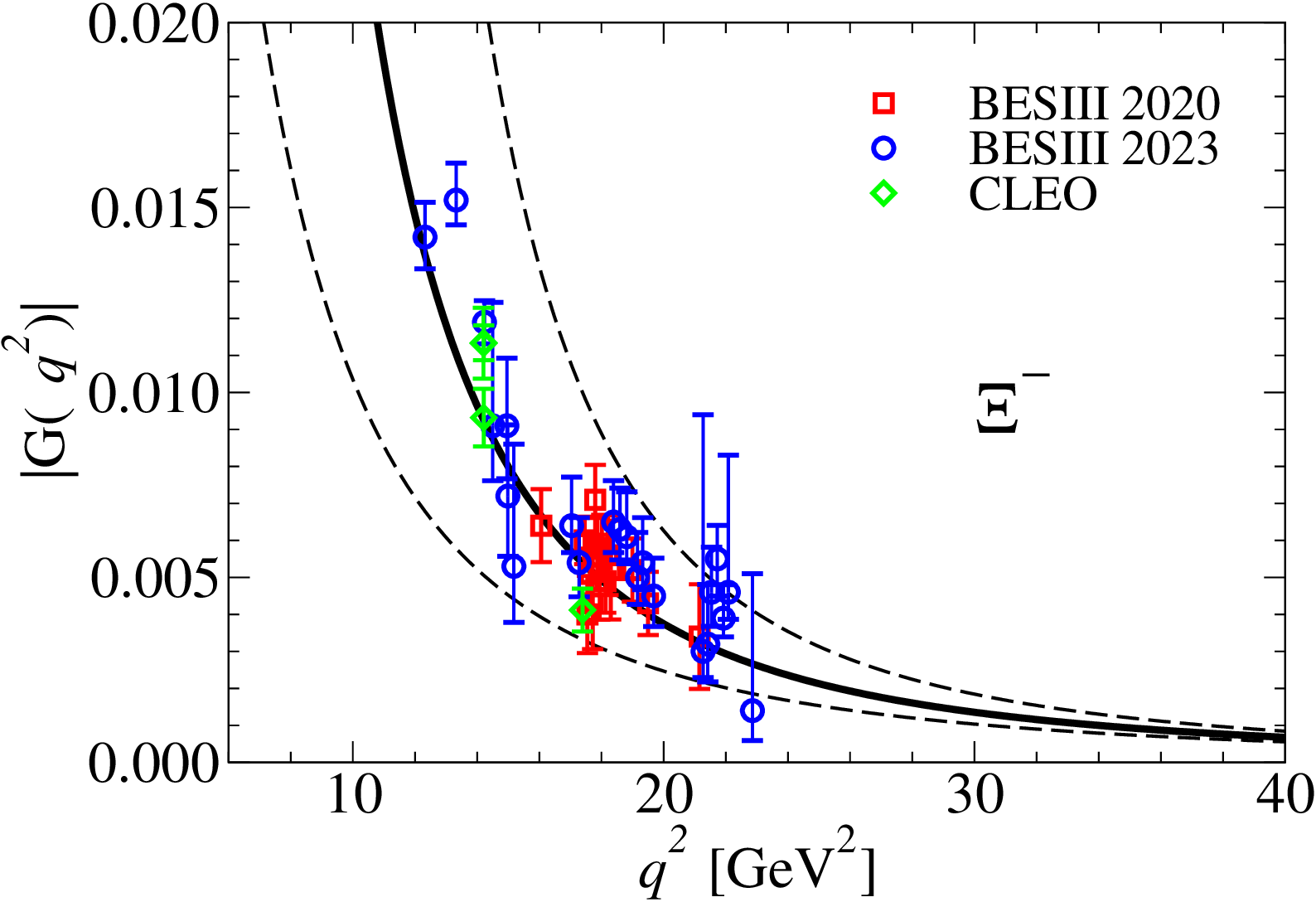} }}
\caption{\footnotesize{Effective form factor $|G(q^2)|$ for $\Xi^-$.
    The data are from CLEO~\cite{Dobbs17a}, BESIII 2020~\cite{BESIII20c} and BESIII 2023~\cite{BESIII23b}.}
  \label{figXi}}

\end{figure*}

We discuss now how we can take into account
the information of the relative angle
between the form factors in the analysis of the ratio $|G_E/G_M|$.
In complete experiments
the ratio  $G_E/G_M$ is determined,
and can be written in the form
\ba
\frac{G_E}{G_M} = \frac{|G_E|}{|G_M|} e^{i \Delta \Phi},
\label{eqGEGMexp}
\ea
where $|G_\ell|$ is the magnitude of
the complex form factor $G_\ell$ and $\Delta \Phi$
is the phase between $G_E$ and $G_M$.
Once the phase is measured,
we can extract the real part of the ratio from
\ba
\left| {\rm Re}\left(\frac{G_E}{G_M} \right) \right|=
\left| \frac{G_E}{G_M} \right| |\cos (\Delta \Phi)|,
\label{eqGEGMexp2}
\ea
which is a fraction of the measured value to $|G_E/G_M|$.
This does not only show that model calculations
of $|G_E/G_M|$ based on real form factors 
agree with the absolute value of the real part of $G_E/G_M$
when the imaginary parts are negligible.
The result (\ref{eqGEGMexp2}) allows also us to infer from the comparison between the
experimental data for the r.h.s.~and our model results
for the ratio $\frac{|G_E|}{|G_M|}$ (where $G_E$ and $G_M$ are real functions)
if the regime where the imaginary parts of
the form factors vanish has been attained. 
The comparison of our model results for $|G_E/G_M|$
with the experimental
values for $|{\rm Re}(G_E/G_M)|$ from Refs.~\cite{BESIII22a,BESIII23c}
that measure phases, are displayed in Fig.~\ref{figure3}.
In the graphs we notice the agreement of our results for the form factors
(considering the model uncertainty  band) and the experimental
data point at the largest $q^2$ point measured,
 $q^2 \simeq 14.3$ GeV$^2$ for the $\Lambda$
and $q^2 \simeq 8.4$ GeV$^2$ for the $\Sigma^+$.
In the graphs we ignore the uncertainties
of the angles $\Delta \Phi$, for simplicity, and used
their central value.
The consideration of the uncertainties on $\Delta \Phi$
only increases the error bars of the $|{\rm Re}(G_E/G_M)|$ data.

It is worth noticing that the agreement between model
calculations and the data is not an indication that
we are in the pQCD regime.
Evidences of scaling with the pQCD behavior
for the nucleon are observed only for $Q^2 \ge 30$ GeV$^2$~\cite{Mergell96a}.
One expects, however, that the convergence for the pQCD regime
can be seen for lower values of $q^2$ for hyperons
due to dynamics of the strange quarks.

Future experiments with more precise data
can help to determine the point where our
model calculations provide a good description of the data
and determine the $q^2$ region where reflection symmetry sets in.


\section{Effective form factors \label{secEFF}}

In a previous work~\cite{HyperonTL} we presented predictions
for the effective form factors (\ref{eqGeff}) for
the baryon octet members $\Lambda$, $\Sigma^{0,\pm}$ and $\Xi^{0,-}$
for $q^2=10$--60 GeV$^2$.
At the time, only $\Lambda$, $\Sigma^+$, $\Sigma^-$, $\Xi^0$ and $\Xi^-$
could be tested in the region of 12--15 GeV$^2$ by the data from CLEO~\cite{Dobbs17a}.
Our estimates were on average consistent with the high $q^2$ data. 
Since then more data for $|G(q^2)|$ became available.
In particular data for $\Sigma^+$ and $\Sigma^-$ became available from BESIII~\cite{BESIII21b}
for the range 5.7--9.1 GeV$^2$.
Surprisingly, the data are in excellent agreement with our estimates 
(namely with the central line values, see Supplementary Material~\cite{SuppM})
even outside the region where model validity was expected to be justified.
A rough estimate of the lower limit for the application
of our model based on the asymptotic
relations, can be $q^2 \approx 10$ GeV$^2$ or $q^2 \approx 8 M_B^2$.

Recently, data from BESIII became available for
$\Lambda$ up to $q^2= 21$ GeV$^2$~\cite{BESIII21a},
and for $\Xi^-$ up to $q^2= 23$ GeV$^2$~\cite{BESIII23a,BESIII21d}.
The new data are compatible with the CLEO data and extend the previous
range covered by the experiments.
The comparison of the new data with the model results
from Ref.~\cite{HyperonTL} is presented in
Figs.~\ref{figLambda2} and \ref{figXi} in a linear scale and in the logarithm scale for $|G(q^2)|$.

It is clear from the figures that there is an excellent agreement between
our predictions and the data for $q^2 > 15$ GeV$^2$, indicating that
we are already in the range of the asymptotic form for the effective form factors.
Compared to the convergence to the pQCD regime, a faster convergence
of the reflection symmetry relations to their asymptotic behavior 
can be expected because in the effective form factors 
$G_M$ dominates on $|G|$ for large $q^2$:
for instance, for the case of the nucleon, $G_M$ dominates over $G_E$,
and $G_M$ converges to the expected power law faster than $G_E$,
in the spacelike region~\cite{Nucleon,Octet2}.

Future data may confirm or deny if we are also close to the
range of the asymptotic region for the ratio $|G_E/G_M|$.


Calculations of the effective form factors
for all the hyperons of the baryon octet are presented in Supplementary Material~\cite{SuppM}
and compared with the world data.

\section{Outlook and conclusions \label{secConclusions}}

Experiments in new facilities (BaBar, CLEO, Belle, PANDA
etc.)~allow us to access
the electromagnetic structure of hyperons in the elastic
timelike region $(q^2 \ge 4 M_B^2$).
Till recently we had access only to
the integrated cross section which can be used
to determine the effective form factor $|G|$,
and encode only the information
of a certain combination of the electric and magnetic
form factors $|G_E|$ and $|G_M|$, 
for spin $\frac{1}{2}$ hyperons.
No separated information about  $|G_E|$ and $|G_M|$
was possible till a few years ago.
Even the separation between $|G_E|$ and $|G_M|$
for the proton and neutron was difficult
and affected by significant errors.

In the last few years it became possible
to determine  $|G_E|$ and $|G_M|$ and the relative phase
between them for $\Lambda$ and $\Sigma^+$.
We expect that the ratio $|G_E/G_M|$ became
accessible in the following years for most
ground state hyperons ($\Sigma^{0,-}$ and $\Xi^{0,-}$),
as well as for charmed baryons (starting with $\Lambda_c^+$).
Our calculations show the start of a partial 
agreement with the data for $q^2 \simeq 14$ GeV$^2$ ($\sqrt{s} \simeq 3.8$ GeV),
which may be an indication that our model
calculations may be compared with future measurements
in the range of $q^2 =20$--30 GeV$^2$
($\sqrt{s}= 4.5$--5.5 GeV) or higher.
Also the separation between the real and imaginary
components may be tested in that range of energies.

\section*{Acknowledgments}
G.R.~was supported the Basic Science Research Program 
through the National Research Foundation of Korea (NRF)
funded by the Ministry of Education  (Grant No.~NRF–2021R1A6A1A03043957).
M.T.P~was supported by the Portuguese Science Foundation FCT
under project CERN/FIS-PAR/0023/2021, and FCT computing project 2021.09667.
K.T.~was supported by Conselho Nacional de Desenvolvimento 
Cient\'{i}fico e Tecnol\'ogico (CNPq, Brazil), Processes No.~313063/2018-4 
and No.~426150/2018-0, and FAPESP Process No.~2019/00763-0
and No.~2023/07313-6, and his work was also part of the projects,
Instituto Nacional de Ci\^{e}ncia e 
Tecnologia - Nuclear Physics and Applications 
(INCT-FNA), Brazil, Process No.~464898/2014-5.
The work of M.K.C.~was supported by the National
Research Foundation of Korea
(Grant Nos.~NRF--2021R1A6A1A03043957 and NRF--2020R1A2C3006177).




%
%

\pagebreak
\clearpage 
~\vspace{1cm} 
\parbox{16cm}{
\begin{center}
  \textbf{\large
Electromagnetic $|G_E/G_M|$ ratios of hyperons
at large timelike $q^2$} \\
    {\large Supplementary Material}
\end{center}}
\setcounter{equation}{0}
\setcounter{figure}{0}
\setcounter{table}{0}
\setcounter{page}{1}
\makeatletter
\renewcommand{\theequation}{S\arabic{equation}}
\renewcommand{\thefigure}{S\arabic{figure}}
\renewcommand{\bibnumfmt}[1]{[S#1]}
\renewcommand{\citenumfont}[1]{S#1} 

\begin{center}
\parbox{16cm}{We present here calculations of the form factors
$G_E$ and $G_M$ and the effective form factor $G$
in the timelike region for the hyperons
$\Lambda$, $\Xi^{-,0}$ and  the $\Sigma^{0,\pm}$.}
\end{center}

\vspace{-.5cm}

\begin{figure*}[h]
\centerline{
\mbox{
\includegraphics[width=2.2in]{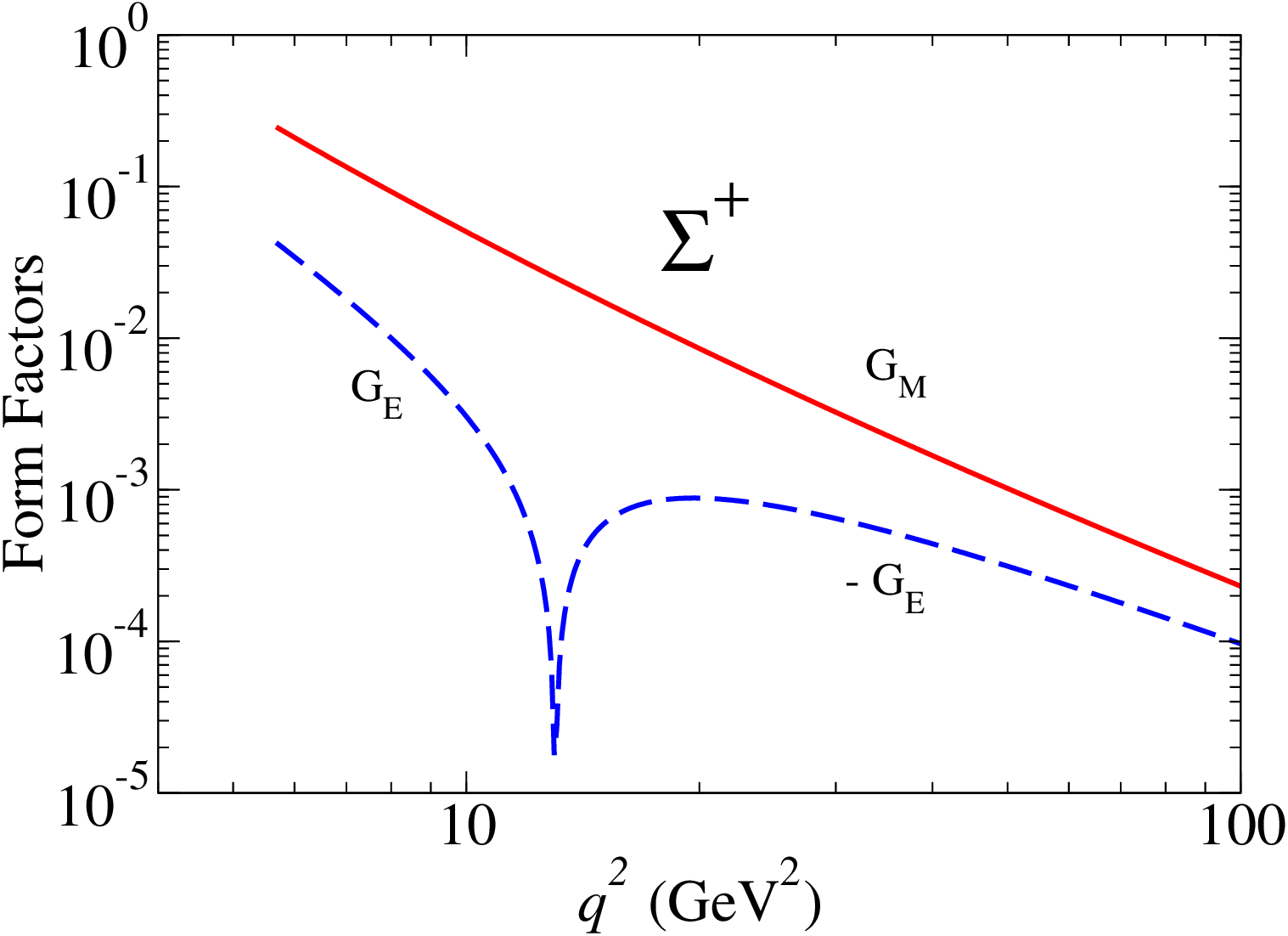} \hspace{.2cm}
\includegraphics[width=2.2in]{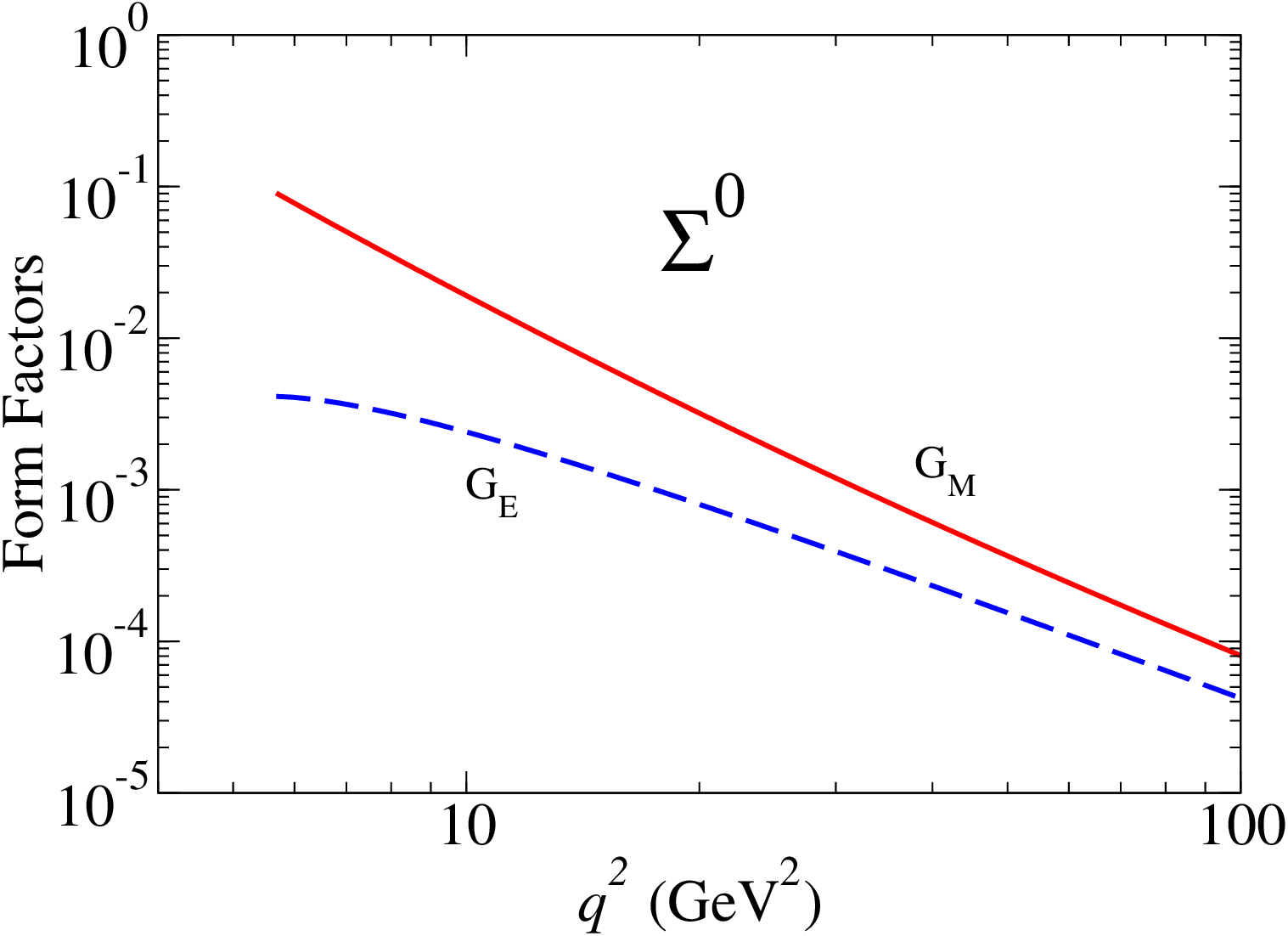} \hspace{.2cm}
\includegraphics[width=2.2in]{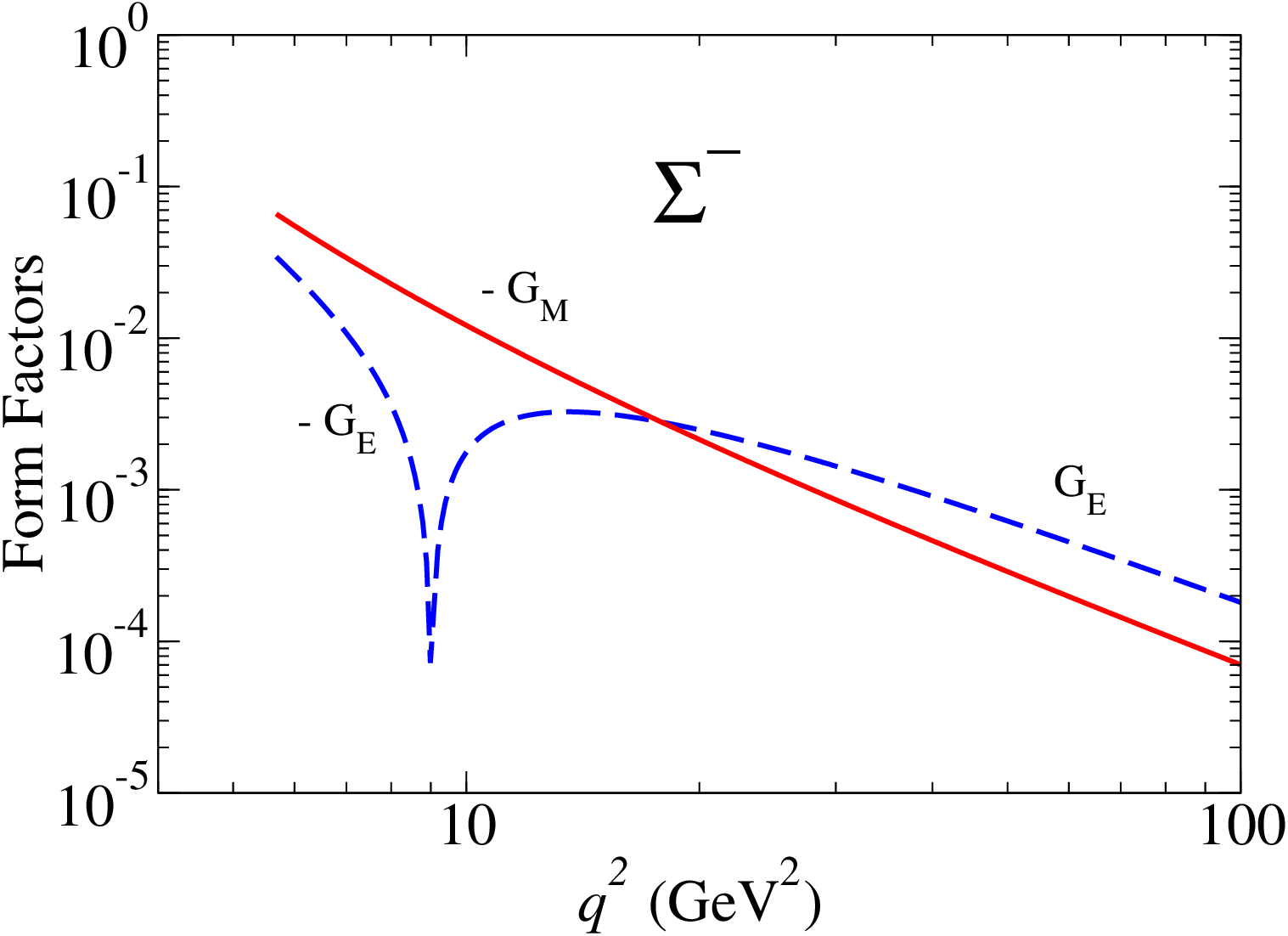}
}}
\centerline{  \vspace{.15cm}}
\centerline{
  \mbox{
\includegraphics[width=2.2in]{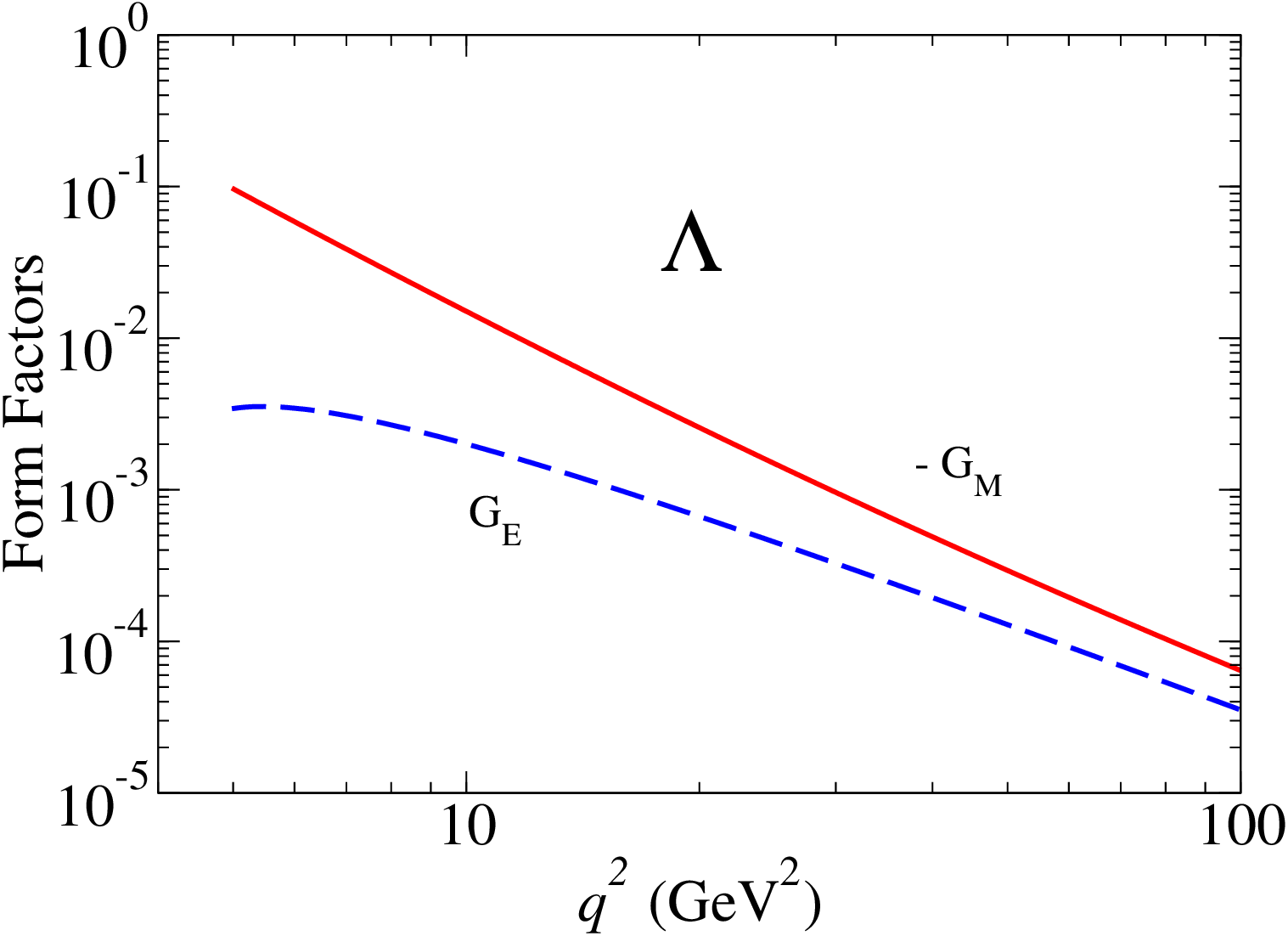} \hspace{.2cm}
\includegraphics[width=2.2in]{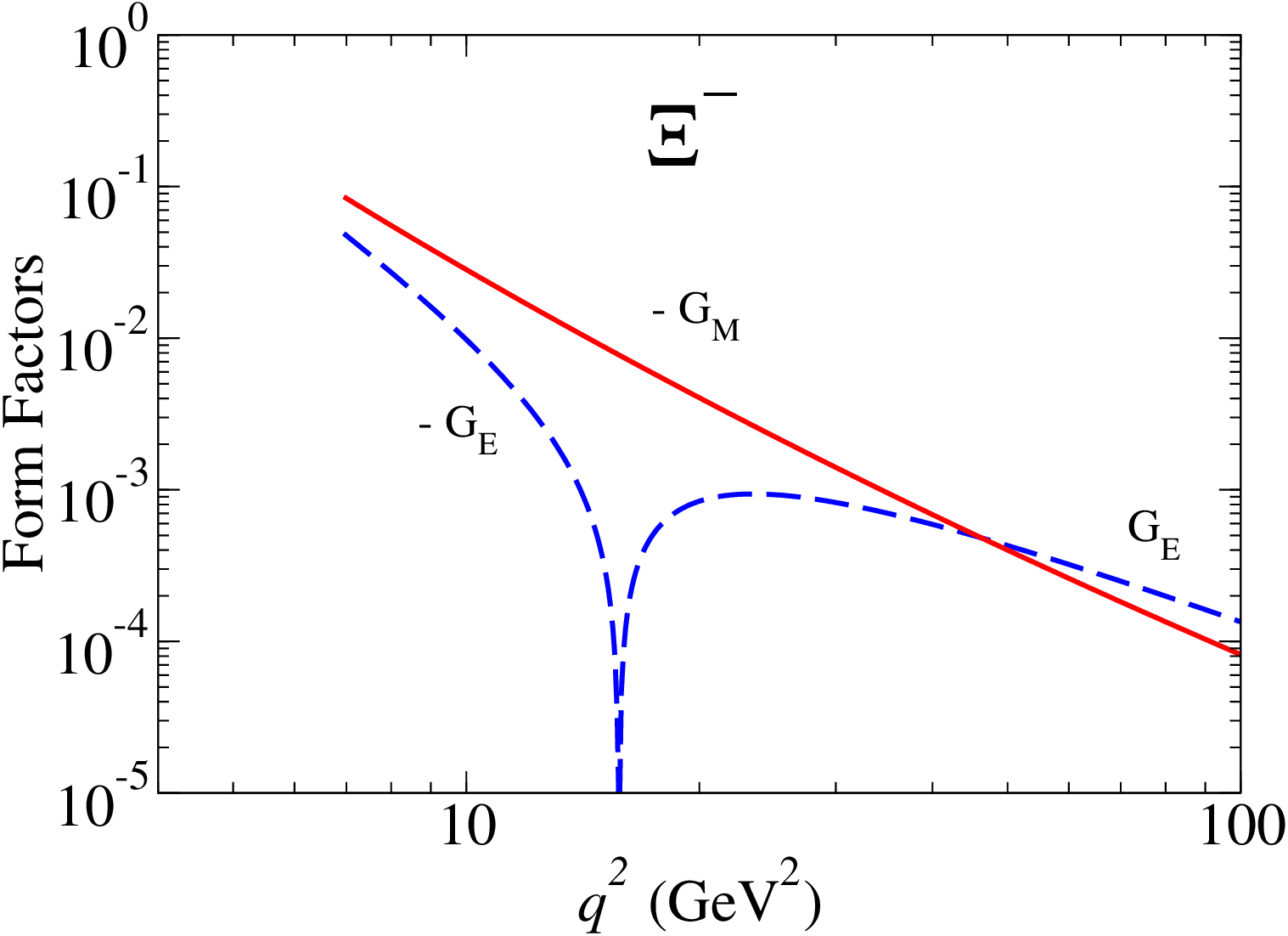} \hspace{.2cm}
\includegraphics[width=2.2in]{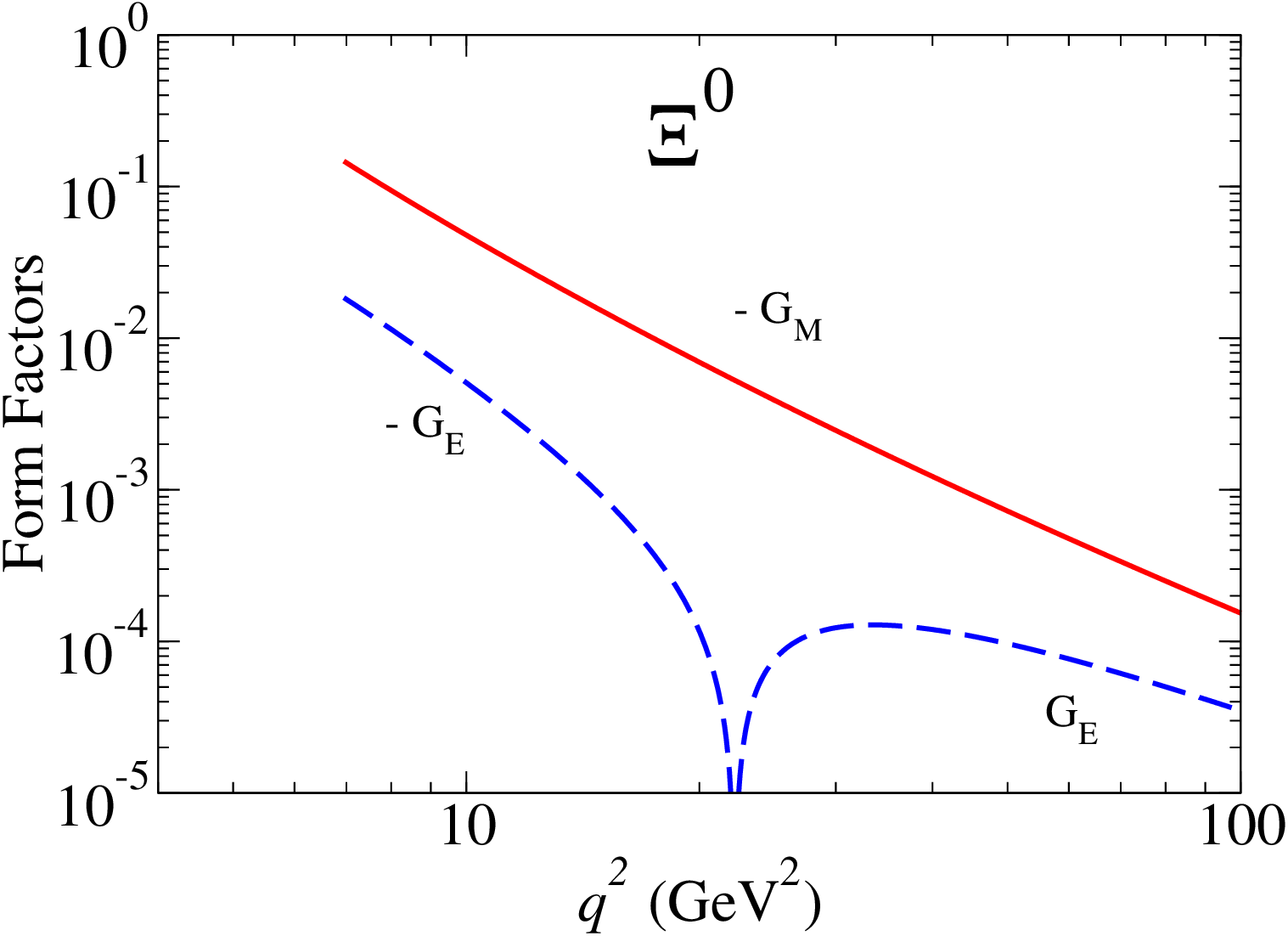}
}}
\caption{\footnotesize{Model calculations of $G_E$ and $G_M$.}}
\label{figure10}
\end{figure*}

\begin{center}
\parbox{16cm}{
  
\section*{Form factors $G_E$ and $G_M$}

The absolute values of the form factors $G_E$ and $G_M$
are presented in Fig.~\ref{figure10}.
The signs of $G_E$ and $G_M$ can be inferred from
the labels included in the figures.
The representation $-G_E$ or $-G_M$ indicate
that the functions are negative in that region.
The cusps indicate the zero of $G_E$.}
\end{center}
\vspace{.25cm}



\begin{figure*}[h]
\centerline{
\mbox{
\includegraphics[width=2.2in]{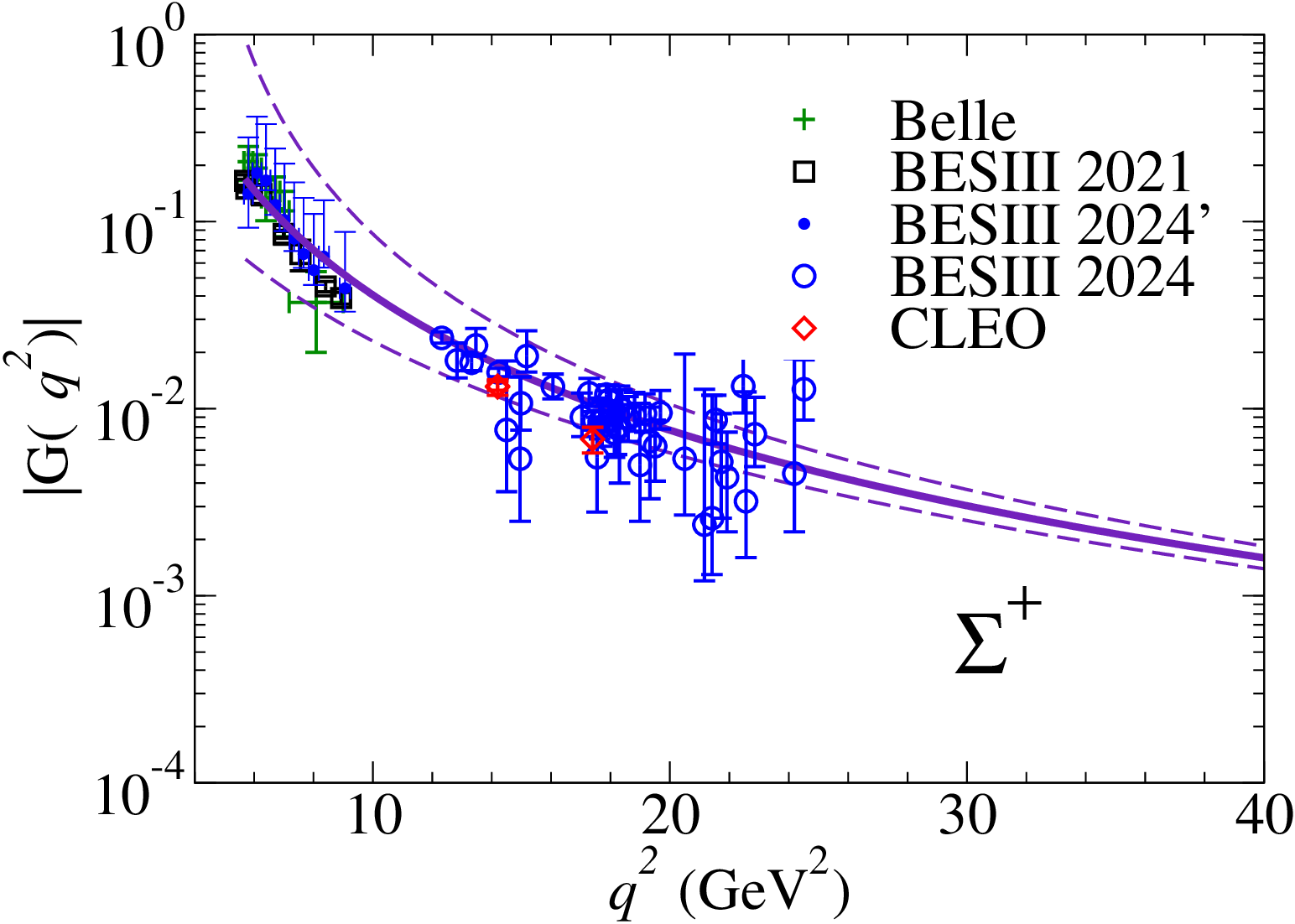} \hspace{.2cm}
\includegraphics[width=2.2in]{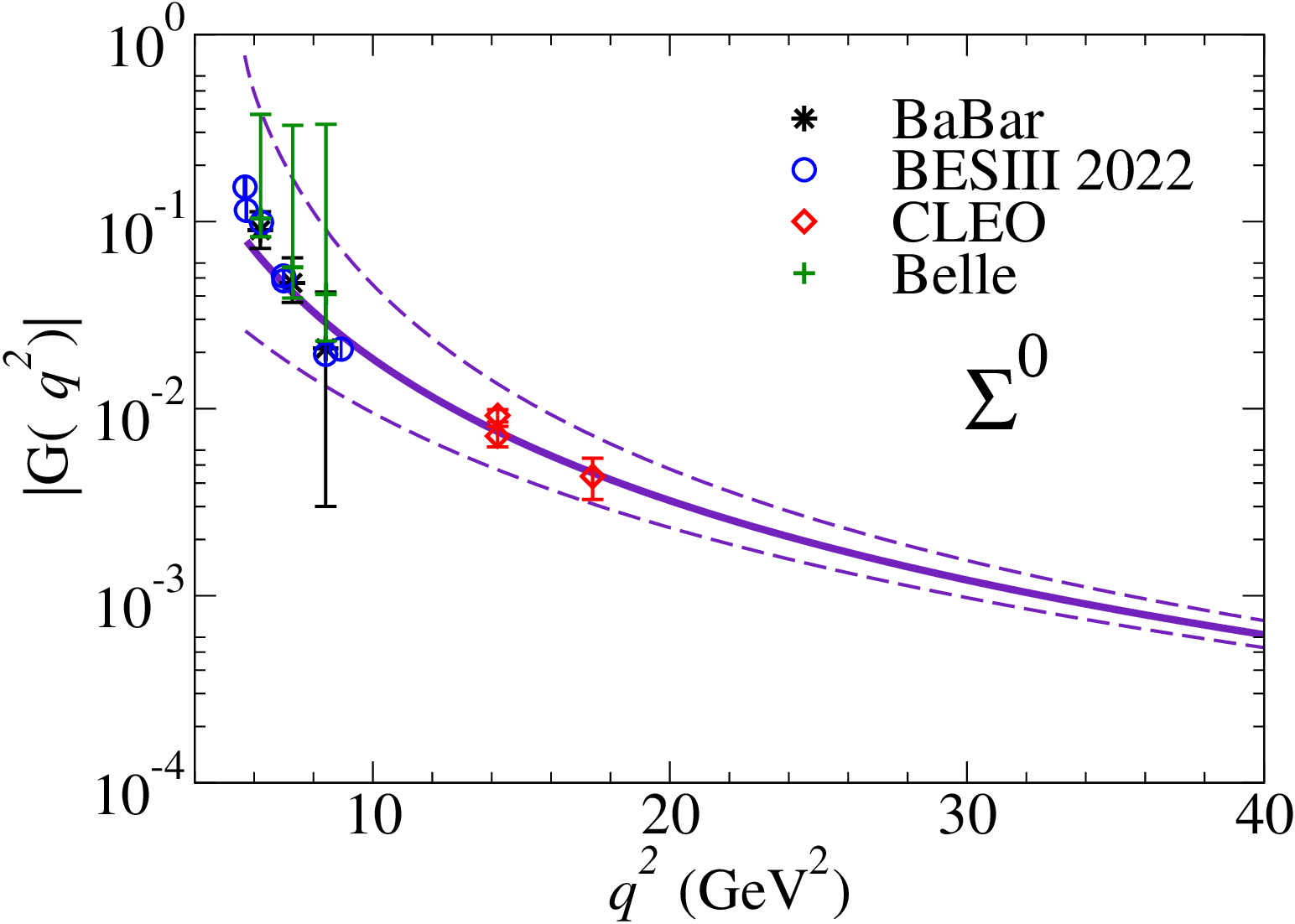} \hspace{.2cm}
\includegraphics[width=2.2in]{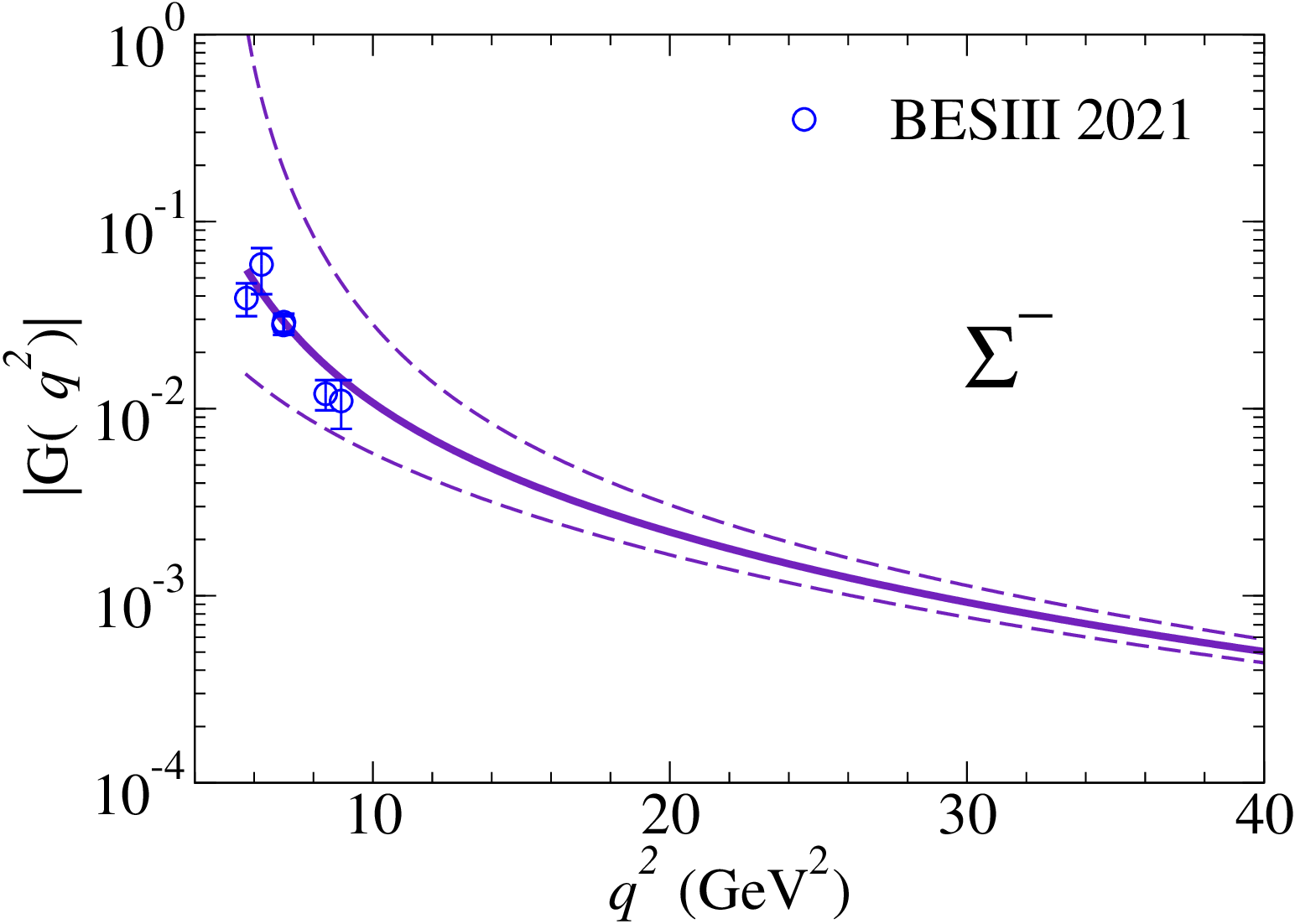} }}
\centerline{  \vspace{.2cm}}
\centerline{
\mbox{
\includegraphics[width=2.2in]{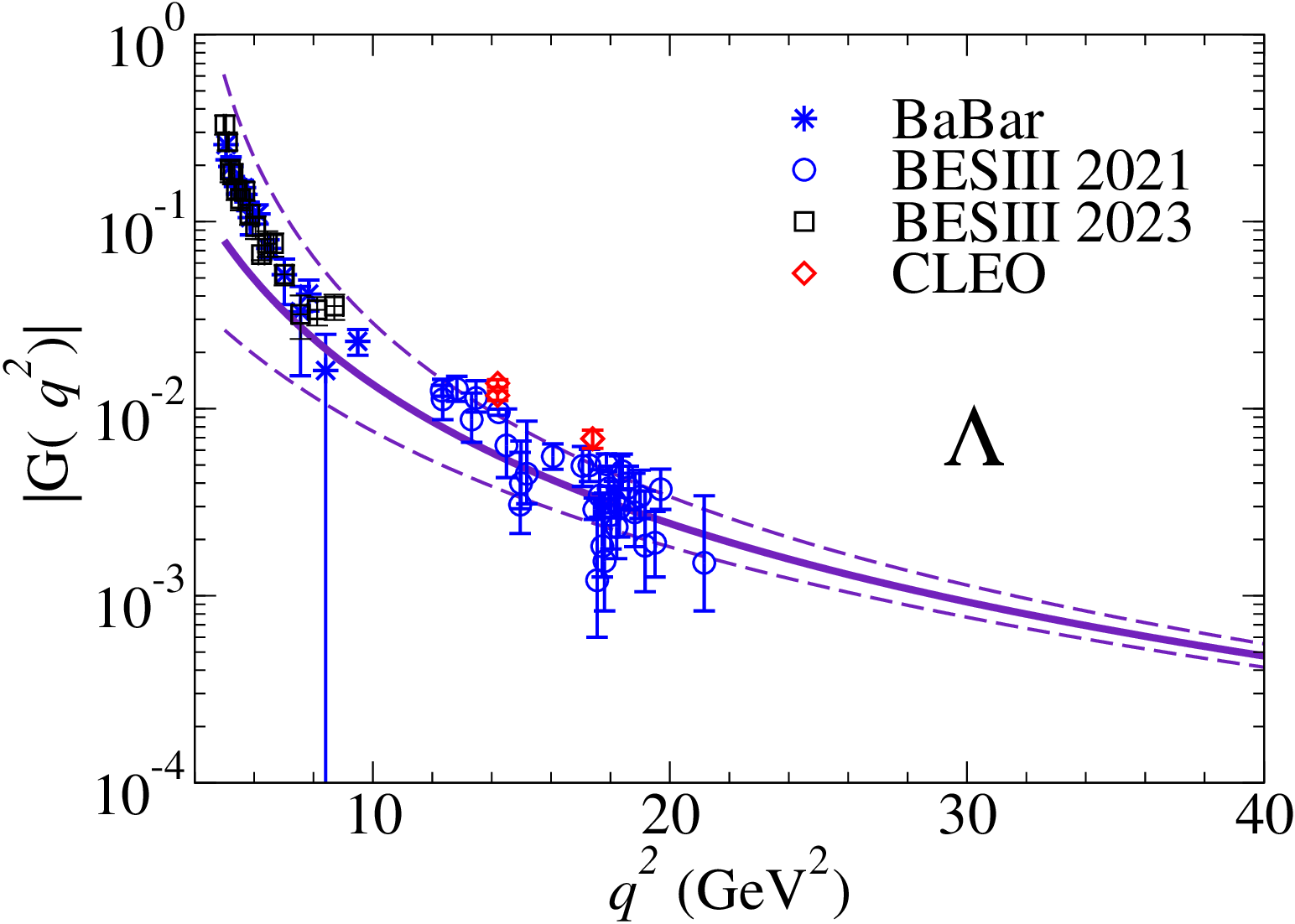} \hspace{.2cm}
\includegraphics[width=2.2in]{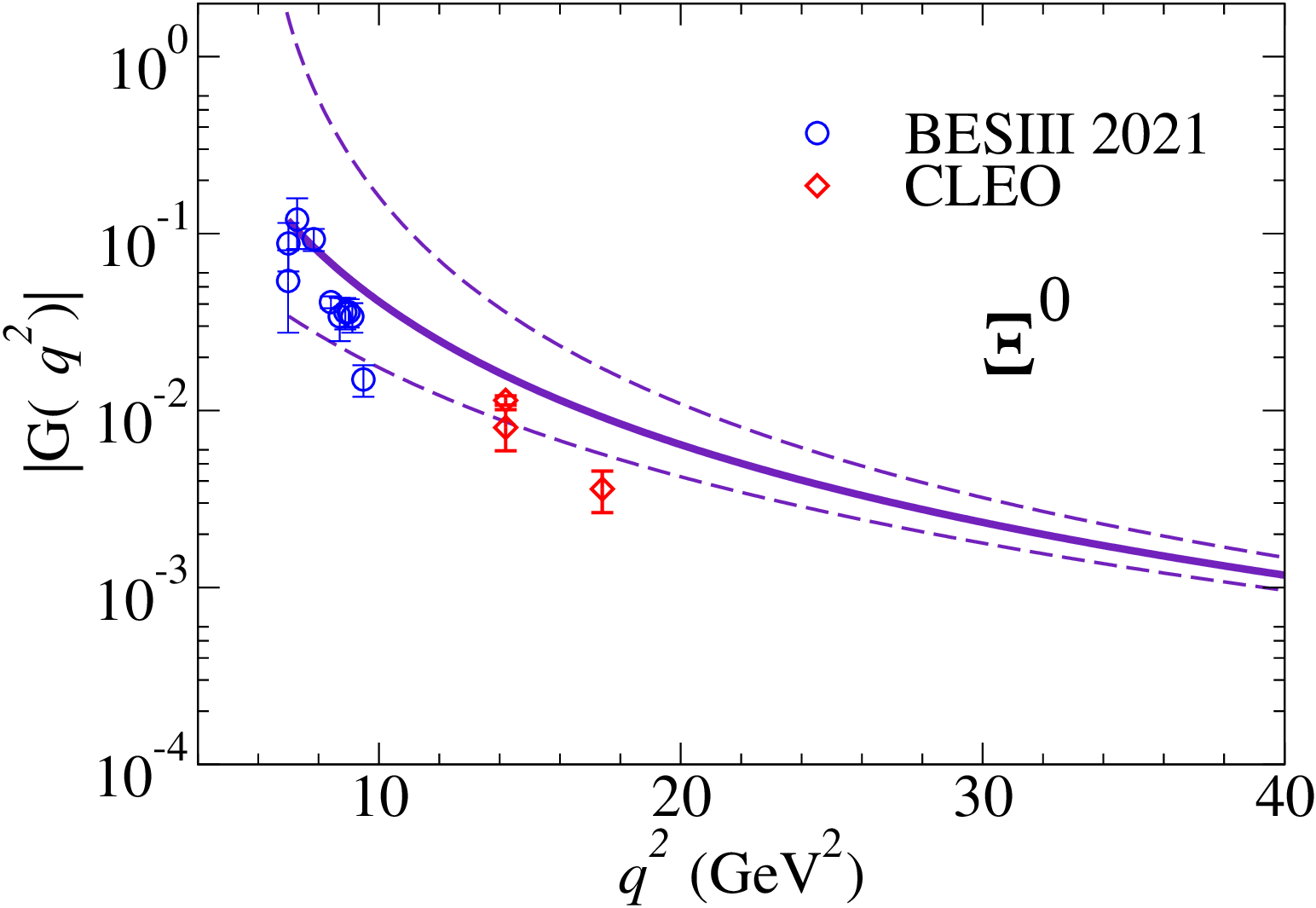} \hspace{.2cm}
\includegraphics[width=2.2in]{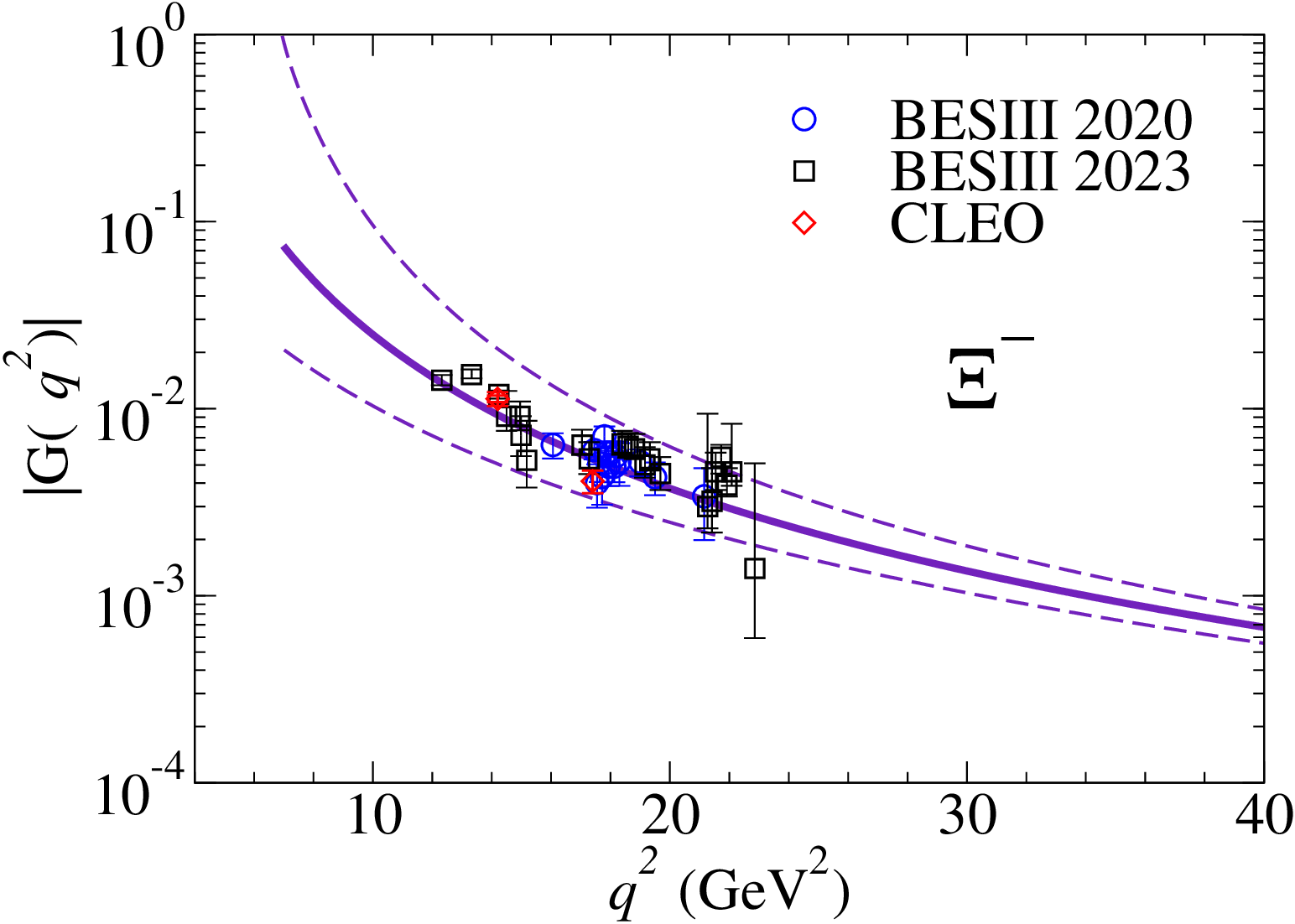}} }
\caption{
\footnotesize{Model calculations of hyperon effective form factors compared with the data.}}
\label{figure20}
\end{figure*}

\begin{center}
\parbox{16cm}{
  \section*{Effective form factors $G$}

The effective form factors are presented in Fig.~\ref{figure20}.
The limits of the theoretical calculations
are represented by the dashed lines.

{\bf Data} $\Sigma^+$:  Belle~\cite{Belle23aX},
BESIII~\cite{BESIII21dX,BESIII24aX,BESIII24bX},
and CLEO~\cite{Dobbs14aX,Dobbs17aX};
{\bf Data} $\Sigma^0$: BaBar~\cite{Aubert07aX}, BESIII~\cite{BESIII22X2}, 
CLEO~\cite{Dobbs14aX,Dobbs17aX} and Belle~\cite{Belle23aX};
{\bf Data} $\Sigma^-$: BESIII~\cite{BESIII21dX}; 
{\bf Data} $\Lambda$: BaBar~\cite{Aubert07aX}, BESIII~\cite{BESIII21aX,BESIII23dX}
and CLEO~\cite{Dobbs14aX,Dobbs17aX};
{\bf Data} $\Xi^0$: BESIII~\cite{BESIII21bX} and CLEO~\cite{Dobbs14aX,Dobbs17aX};
{\bf Data} $\Xi^-$: BESIII~\cite{BESIII20cX,BESIII23bX} and
CLEO~\cite{Dobbs14aX,Dobbs17aX}.}

\end{center}

\vspace{1cm}

\parbox{16cm}{
\begin{center}

\end{center}}


\end{document}